\documentclass[11pt]{article} 
\usepackage{arxiv}

\usepackage{graphicx}
\usepackage{amssymb}
\usepackage{amsmath,stmaryrd} 
\newcommand{\bs}{\boldsymbol}
\newcommand{\eq}{Eq.~}

\newcommand{\fig}{Fig.~}

\usepackage{natbib}

\title{Network-based analysis of fluid flows:\\progress and outlook}
\renewcommand{\shorttitle}{\small Network-based analysis of fluid flows}
\author{Kunihiko Taira$^1$~\& Aditya G. Nair$^2$\\
{$^1$ Department of Mechanical and Aerospace Engineering, University of California, Los Angeles, CA 90095}\\
{$^2$Department of Mechanical Engineering, University of Nevada, Reno, NV 89557}
\\
}
\date{}

\begin{document}

\maketitle

\begin{abstract}
The network of interactions among fluid elements and coherent structures gives rise to the incredibly rich dynamics of vortical flows.  These interactions can be described with the use of mathematical tools from the emerging field of network science, which leverages graph theory, dynamical systems theory, data science, and control theory.  The blending of network science and fluid mechanics facilitates the extraction of the key interactions and communities in terms of vortical elements, modal structures, and particle trajectories.  Phase-space techniques and time-delay embedding enable network-based analysis in terms of visibility, recurrence, and cluster transitions leveraging available time-series measurements.  Equipped with the knowledge of interactions and communities, the network-theoretic approach enables the analysis, modeling, and control of fluid flows, with a particular emphasis on interactive dynamics.  In this article, we provide a brief introduction to network science and an overview of the progress on network-based strategies to study the complex dynamics of fluid flows. Case studies are surveyed to highlight the utility of network-based techniques to tackle a range of problems from fluid mechanics. Towards the end of the paper, we offer an outlook on network-inspired approaches.
\end{abstract}

\tableofcontents


\section{Introduction}

{\it Network} is a familiar word in our conversations to describe a web of connections amongst a group of people or a collection of physical elements.  Social networks, airline networks, power grid networks, communication networks, neural networks, and food chain networks are some of the well-known examples of networks.  
Over the past three decades, network science has emerged as an interdisciplinary field of study, comprised of a range of disciplines, including graph theory, dynamical systems, statistical physics, data science, control theory, and biology \citep{Barabasi:SA03,Newman:PhysicsToday08,Newman18,Barabasi16,Dorogovtsev10,Latora17,Cohen10}.  
The implications of network science research are not only limited to the academic arena but also seen by a much larger audience in the general public.  

As our society becomes increasingly interactive, the transfer of information, natural resources, and people on networks are serving as critical backbones of our daily lives.  The importance of these networks cannot be overstated.  In fact, we have experienced firsthand how dynamics on such networks can be venerable when a virus like COVID-19 spreads through human interactions \citep{gatto2020spread,weitz2020modeling,firth2020using}.  However, we have also seen how resilient human communications are as we swiftly transitioned to a remote working environment on the virtual network.  For these reasons, a glimpse of network science has made its way into the public eyes during the COVID-19 pandemic. Some of the terms, such as clusters, hubs, and pandemic modeling from network science have become part of the household lexicon.

The origin of network science is deeply rooted in graph theory dating back to Leonhard Euler's solution to the K\"{o}nigsberg bridge problem \citep{Euler:bridge}. Mathematical studies of a group of interconnected elements led to the use of techniques from algebra and geometry giving rise to graph theory \citep{Bollobas98,Diestel17}. More recently, there has been a renewed interest led by those originally in statistical physics to lay the foundation to analyze large-scale modern networks \citep{Newman:SIAMReview03,NS_Nat_Academy}. Their efforts have been widely supported by the industry and government to study communication networks \citep{faloutsos2011power}, information networks \citep{sun2013mining}, and social networks \citep{otte2002social}, with implications for resource management \citep{prell2009stakeholder} and national security \citep{ressler2006social}. More broadly, applications have been seen in semantic networks in linguistics \citep{sowa1987semantic}, ecological networks \citep{montoya2006ecological}, biological networks \citep{junker2011analysis}, and knot theory in algebraic topology \citep{crowell2012introduction}. One of the successful applications of network science is in mapping the structural and functional connections in the human brain \citep{Bassett:NS06,Lynn:NRP19}.  It is now expected that these studies will support medical research aimed to treat conditions, including Alzheimer's disease and schizophrenia \citep{stam2014modern}.
Moreover, developing strategies to distribute a limited supply of vaccinations to a large population during the age of pandemic has benefited from network-based transmission models.

Physical interactions are ubiquitous in continuum mechanics.
In fluid mechanics, the presence of interactions among a group of vortices and flow structures give rise to complex dynamics, spatio-temporal multiscale properties, and turbulence \citep{Taira2016jfm}. These interactions have traditionally been studied through painstaking theoretical endeavors. 
Network science may provide the general framework to extract the underlying structure of interactions, regardless of the particular flow problem of interest. 
In fact, one of the unique capabilities of network science is to decompose the structural and dynamical properties of complex interactions.
The time is now ripe for fluid mechanics to consider network science-inspired analysis because of advanced large-scale network algorithms and increased resources, as discussed by a recent survey paper \citep{Iacobello:PA21}. 

The purpose of this paper is to provide an introduction to network science and an extensive overview of the current status of the use of network analysis in fluid mechanics.  
Although both of these fields are vast in their coverage, the overlap of these disciplines has only emerged over the past few years. 
Let us note that the properties of interactions commonly studied in network science and fluid mechanics are vastly different.  
For this reason, applications of network science to fluid mechanics require care and innovation to establish an amicable framework that extends the horizon of studying flow field interactions.

In this survey, we present some of the latest progress in analyzing some of the most challenging problems in fluid mechanics through the lens of network science. The interactions present in fluid flows give rise to the rich complex dynamics observed in unsteady wakes, turbulence, transition, combustion, and flow control.
The fundamental interactions amongst a collection of vortices have traditionally been studied through the Biot-Savart law. 
Recently, a network-inspired analysis has been presented to model the interactive dynamics of the vortices \citep{Nair:JFM15}. This new formulation has been applied to analyze inviscid flows \citep{Nair:JFM15}, viscous wake flows \citep{meena2018network}, isotropic turbulence \citep{Taira2016jfm,MGM:JFM21} and combustion \citep{Murayama:PRE18, sujith2020complex}.  
Characterizing the modal interaction in wave-space, there have been emerging studies on examining energy transfers in wake dynamics \citep{nair2018network} and turbulence cascades \citep{gurcan2020turbulence}. Lagrangian perspectives can also be taken to formulate particle proximity networks that can detect transitions features in vortical and turbulent flows \citep{schlueter2017jfm,iacobello2019lagrangian}. 

The above networks have been founded with spatial information. Fluid flow networks can also be formulated from sensor-based measurements through visibility graphs \citep{lacasa2012time}, recurrence networks \citep{donner2011recurrence} and cluster-based networks \citep{Kaiser:JFM14}. Visibility graphs and recurrence networks have been used as a diagnostic tool to discover order from what may appear as chaotic data \citep{lacasa2008time}. These tools have been used to observe transitional dynamics in turbulence \citep{iacobello2018visibility}, geophysical flows \citep{donner2012visibility} and chemically reacting flows \citep{gotoda2017characterization, godavarthi2017recurrence}. In particular, extracting network scale-invariance was found to be an indicator for detecting the onset of violent instabilities in combustion systems \citep{murugesan2015combustion}.

Fluid dynamics have unique challenges in quantifying physical interactions, compared to what is ordinarily studied in network science. Fluid phenomena occur over a continuous spectrum of scales yielding dense and weighted interactions. This is in contrast to social networks where the network is often sparse and unweighted. The current toolsets from network science may require extensions to adapt them to fluid mechanics applications. Furthermore, the high degrees of freedom required to describe fluid flow physics may call for an increase in computational requirements. Nonetheless, the current explosion of ideas from data science \citep{Bishop06,Brunton2019book,Watt20,Goodfellow16} provides a unique opportunity to extend the network-inspired analysis to examine the dynamics and structures of fluid flows. We believe the time is ripe to assemble the collection of works in the community that focus on studying interactions in fluid flows from a network-centric perspective.

The objective of this survey paper is to stimulate the use of network-inspired techniques to study fluid flows with complex dynamics and structures.  In section \ref{sec:networks}, we provide a brief introduction to network science and its fundamental analysis techniques.  We present the definition of a network and explain how important nodes and structures can be extracted.  We further discuss how dynamics on networks can be examined, modeled, and inferred.  In section \ref{sec:flows}, network-based studies of fluid flows are reviewed.  The network formulations in these studies are broadly categorized into {\it flow field networks} and {\it sensor networks}.  The former approach incorporates spatial information into the definition of the network, and the latter formulation is based on sensor time-series.  A variety of fluid flow problems are presented in this section.  In section \ref{sec:outlook}, we provide some outlook on the use of network-based techniques and discuss some of the challenges.  Finally, in section \ref{sec:conclusions}, we end the paper with concluding remarks.


\section{Networks}
\label{sec:networks}

The field of network science is vast and rapidly growing.  In this section, we offer a brief introduction to networks and relevant analysis techniques that may be useful for fluid dynamics.  The materials contained herein are focused on weighted networks as fluid flows possess a broad range of scales in time and space.  Readers interested in additional details and materials are directed to a collection of network science review articles \citep{Newman:PhysicsToday08, Newman:SIAMReview03, Albert:RMP02} and textbooks \citep{Newman18, Dorogovtsev10, Estrada12, Barabasi16, Latora17, Mesbahi10}.

\subsection{Fundamentals}

Let us consider a collection of elements that are interconnected.  We refer to these elements as {\it nodes} (or {\it vertices}) and denote them as $v_i$.  If nodes $v_i$ and $v_j$ are connected ({\it adjacent}), we say that there is an {\it edge} or a {\it link} between them and express it as $e_{ij}$ (also denoted as $v_i v_j$ or $(i,j)$).  A {\it network} is a connected structure comprised of a set of vertices (nodes) $\mathcal{V} = \{ v_1, v_2, \dots, v_n\}$ and a set of edges (links) $\mathcal{E}$.  Mathematically, the network is defined by a {\it graph} 
\begin{equation}
	\mathcal{G} = \mathcal{G}(\mathcal{V},\mathcal{E}).
\end{equation}  
This type of graph is known as an unweighted graph.  

A network can also carry edge weights $w_{ij}$ on edges $e_{ij}$ to constitute a weighted network (graph)
\begin{equation}
	\mathcal{G} = \mathcal{G}(\mathcal{V},\mathcal{E},\mathcal{W}),
\end{equation} 
where $\mathcal{W}$ represents the collection of edge weights.  Here, $w_{ij}$ denotes the connection from $j$ to $i$.  These weights can be symmetric $w_{ij} = w_{ji}$ for undirected networks or can be asymmetric $w_{ij} \neq w_{ji}$ for directed networks. 
 
The graph $\mathcal{G}$ with $n$ nodes can be represented elegantly through an {\it adjacency matrix} $\boldsymbol{A} \in \mathtt{R}^{n \times n}$, where
\begin{equation}
	A_{ij} = 
	\begin{cases} w_{ij} & \text{if edge}~e_{ij} \in \mathcal{E} \\
	0 & \text{otherwise}
	\end{cases}
\end{equation}
If $\mathcal{G}$ is unweighted, the weights $w_{ij}$ take $1$ or $0$.   With the mathematical abstraction of a graph represented in terms of a matrix, we can now use linear algebra to analyze networks.  This adjacency matrix serves as the foundation for graph-theoretic analysis of networks.

\begin{figure*}[ht!]
\centering
	\includegraphics[width=0.38\textwidth]{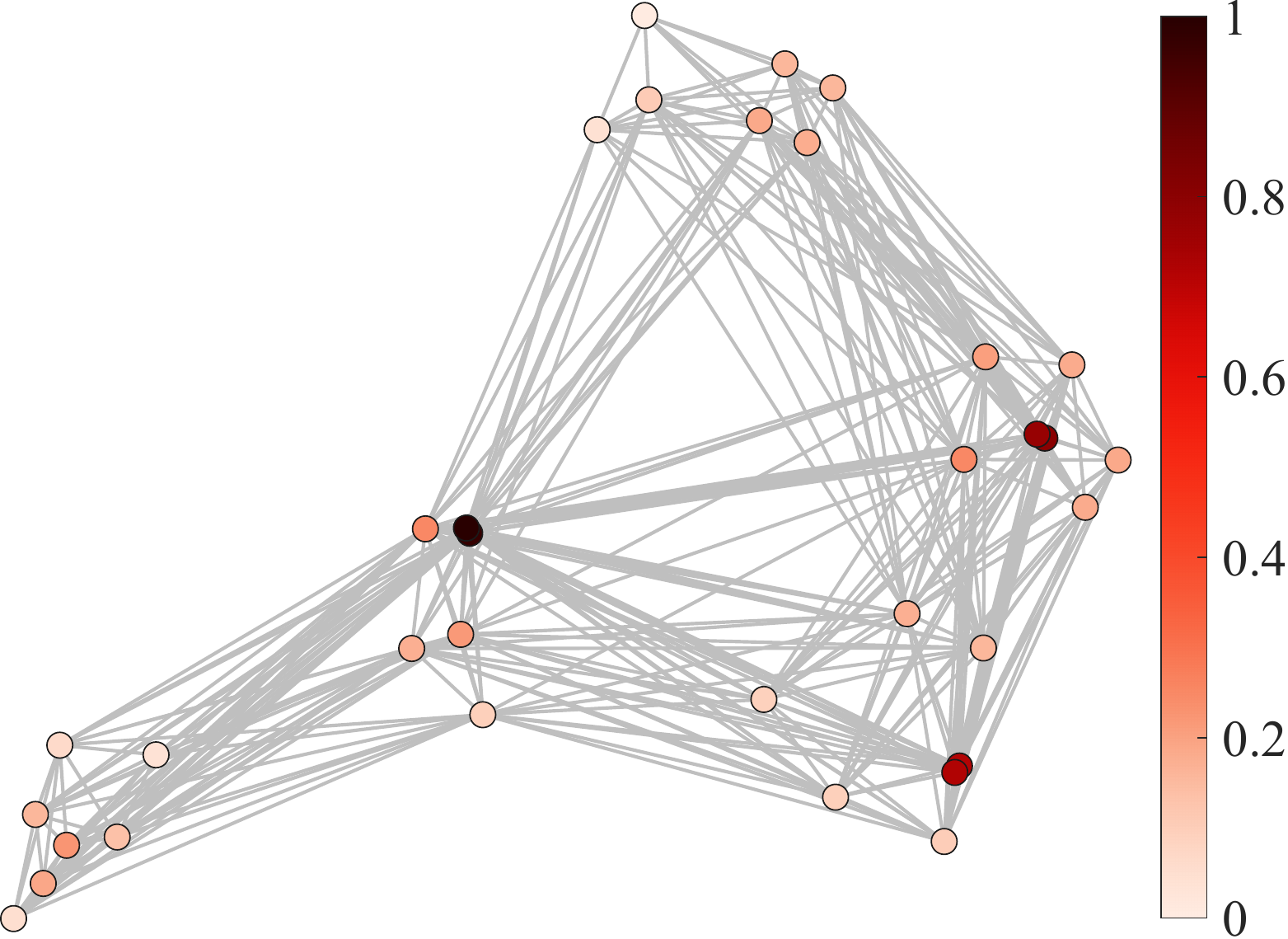}~~~~~
	\includegraphics[width=0.32\textwidth]{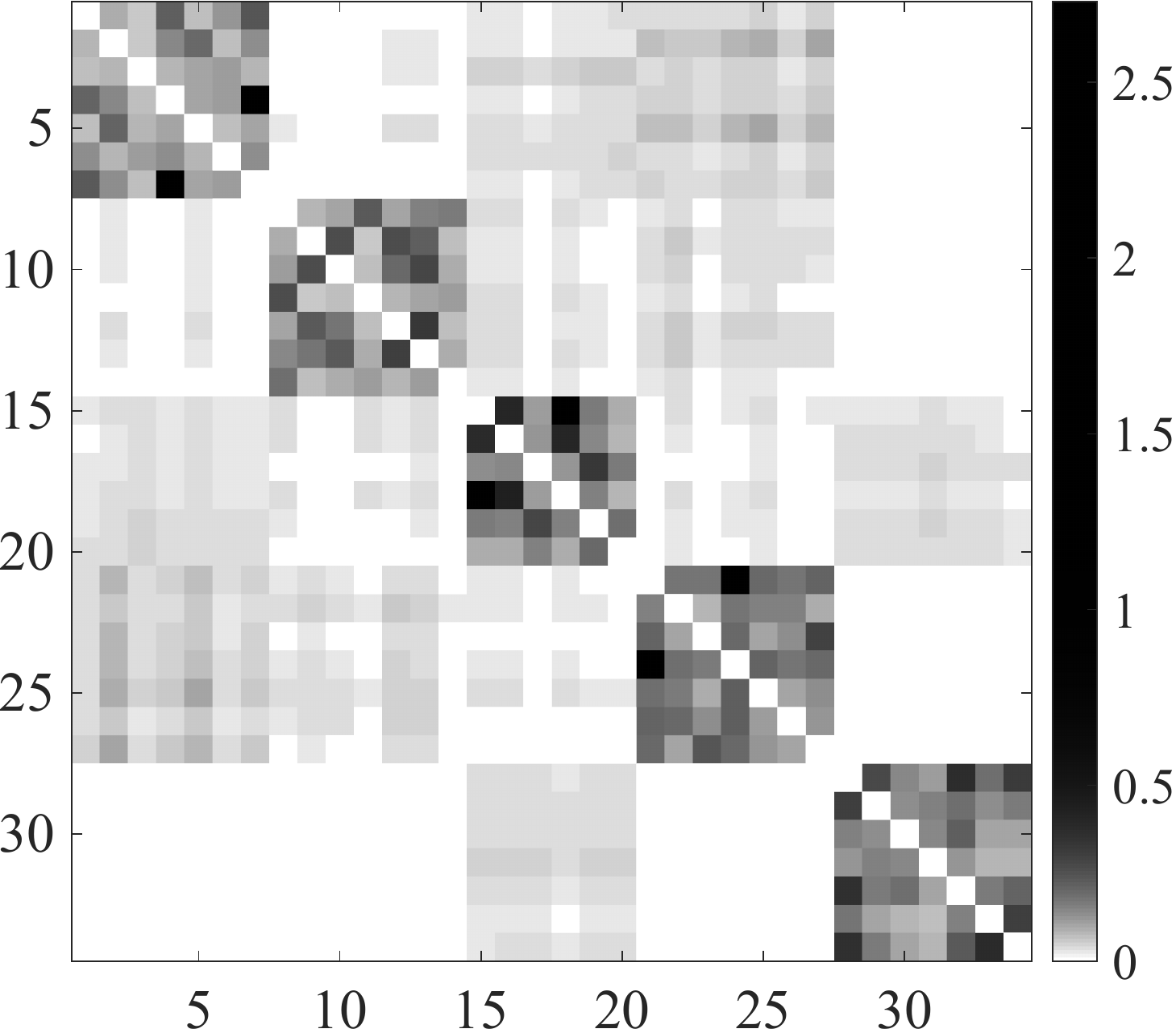}
	\caption{An example weighted vortical-interaction network (left) and the visualization of its adjacency matrix $\boldsymbol{A}$ (right).  The network nodes are colored by the strength distribution $\boldsymbol{s}$.}
	\label{fig:example}
\end{figure*}

To illustrate the concept of a network, let us present an example network, as shown in Figure \ref{fig:example}.  Here, the network nodes are shown in circles and the edges are depicted with lines between the nodes.  This network is provided with symmetric weights, which are visualized with the adjacency matrix in Figure \ref{fig:example}(b).

Given the adjacency matrix $\boldsymbol{A}$, we can determine the strength $\boldsymbol{s}$ of the nodes in terms of their connectivities.  For directed networks, in-strength and out-strength are respectively defined as
\begin{equation}
    s_i^\text{in} = \sum_j A_{ij}, 
    \quad
    s_j^\text{out} = \sum_i A_{ij},
    \label{eq:strength_dir}
\end{equation}
which quantify the inward and outward connectivities of nodes.  For undirected networks, the in-strength and the out-strength are equal to each other ($\boldsymbol{s}^\text{in} = \boldsymbol{s}^\text{out}$).  The strengths are referred to as degrees ($\boldsymbol{k}^\text{in}$ and $\boldsymbol{k}^\text{out}$) for unweighted graphs.   

We can further analyze dynamics on networks with the established framework.  Consider a variable $\varphi_i$ to be defined on the nodes and let it diffuse over a network \citep{Newman18}.  The rate of diffusion from $v_j$ to $v_i$ is the product of the diffusivity $\kappa$ and the difference between the variables, i.e., $\kappa (\varphi_j - \varphi_i)$.  Because this diffusive process takes place only when nodes are adjacent to one another, we can incorporate the adjacency matrix to describe the dynamics
\begin{equation}
    \frac{d \varphi_i}{dt} = \kappa \sum_j A_{ij} (\varphi_j - \varphi_i).
    \label{eq:diffusion}
\end{equation}
Expressing the state variable as $\boldsymbol{\varphi} = (\varphi_1, \varphi_2, \dots, \varphi_n)^T$, we can rewrite the above equation as
\begin{equation}
    \frac{d\boldsymbol{\varphi}}{dt} 
    = \kappa (\bs{A} - \bs{D}) \boldsymbol{\varphi},
    \label{eq:diffusion_vec}
\end{equation}
where matrix $\bs{D} \equiv \text{diag}(\bs{s}^\text{in})$.  This equation describes the diffusion process on a network.  The matrix that appears on right-hand side of this equation
\begin{equation}
    \bs{L} \equiv \bs{D} - \bs{A} \in \mathbb{R}^{n \times n}
\end{equation}
is called the {\it graph Laplacian}, which is another fundamental matrix in graph theory.  
The solution to \eq (\ref{eq:diffusion_vec}) is 
\begin{equation}
    \boldsymbol{\varphi}(t) = \exp( -\kappa \bs{L} t ) \boldsymbol{\varphi}(t_0).
\end{equation}
Note that the graph Laplacian is the analog of the continuous Laplacian $\nabla^2$ but with a sign difference.  
Together with the adjacency matrix, the graph Laplacian serves as the backbone to analyze structures of networks.  Both of these matrices are founded solely on the structural information of the nodal connectivity.

\subsection{Centrality Measures}
\label{centralitymeasures}

It is often useful to identify important nodes on a network from a standpoint of analysis, modeling, and control of networked dynamics.  A measure of connectivity is referred to as a {\it centrality measure}, and there are several measures used to characterize the importance of nodes on a network.

\begin{figure*}[ht!]
\centering
	\includegraphics[width=0.96\textwidth]{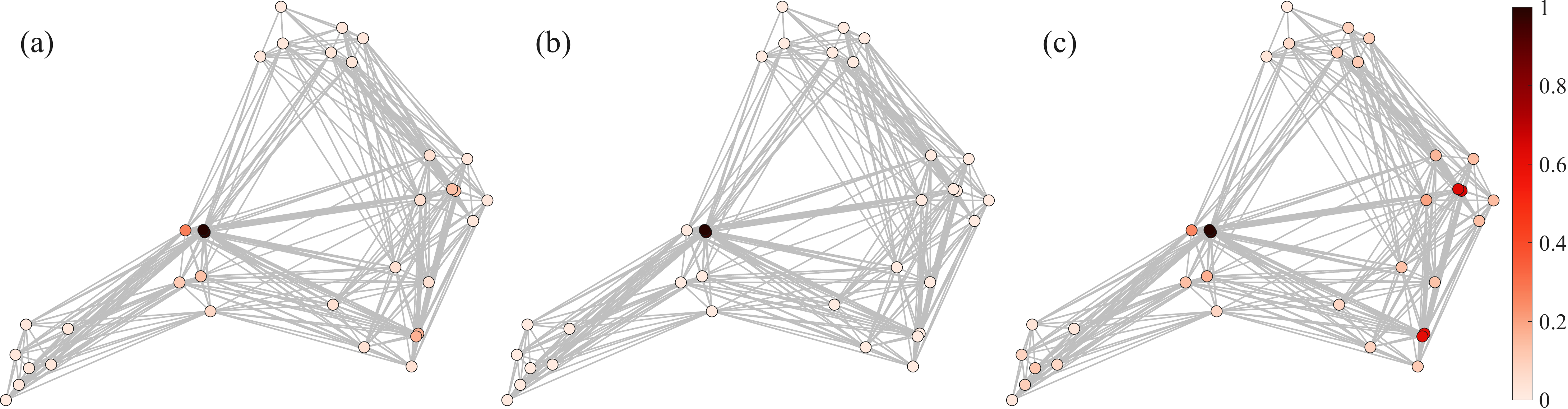}	\caption{Centrality measures for example network in \fig \ref{fig:example}.  (a) Eigenvector centrality of the adjacency matrix.  (b) Eigenvector centrality of the graph Laplacian matrix.  (c) Katz centrality with $\alpha = 1/(2\lambda_1(\boldsymbol{A}))$.}
	\label{fig:strength}
\end{figure*}

\subsubsection*{Strength (degree) centrality}

The most basic centrality measure is the {\it strength centrality} $\boldsymbol{s}$ defined in Eq.~(\ref{eq:strength_dir}).  Since the nodal strength is referred to as the nodal degree for unweighted networks, this centrality measure is also known as the {\it degree centrality}.  The effect of outward connections from a node is described by the out-degree $\boldsymbol{s}^\text{out}$ and the strength of incoming connections into a node is described by the in-degree $\boldsymbol{s}^\text{in}$ from Eq.~(\ref{eq:strength_dir}).  Nodes with high out-degree values are considered to have a strong influence over the network.  On the other hand, nodes with high in-degree can be viewed as being strongly influenced or popular.  

The strength centrality is visualized in Figure \ref{fig:example} for the previously shown undirected network example.  Here, we observe that there are three nodes (in black and dark red) particularly, with high strength centrality values.  These nodes possess a collection of edges that have high edge weights.  It can also be noticed that these nodes act as central nodes within their subregions of the overall network.  

\subsubsection*{Eigenvector centrality}

In many instances, we are concerned with the dynamics on a network.  The simplest dynamics on a network would be to transfer a variable of interest $\varphi$ from a node to its adjacent (connected) neighbors through 
\begin{equation}
	\varphi_i^{m+1} = \sum_j A_{ij} \varphi_j^m.
	\label{eq:Ax}
\end{equation} 
For this type of dynamics, it is often desirable to determine the dominant pattern for the transfer of variable $\boldsymbol{\varphi}$.  This can be revealed by considering the eigenvalue problem of
\begin{equation}
	\boldsymbol{A} \boldsymbol{v} = \lambda \boldsymbol{v},
\end{equation} 
where $\boldsymbol{v}$ is an eigenvector and $\lambda$ is its corresponding eigenvalue.
For a symmetric network, eigenvalues of $\boldsymbol{A}$ are all real.

The leading eigenvector $\boldsymbol{v}_1$ of the adjacency matrix $\boldsymbol{A}$ (associated with the largest eigenvalue $\lambda_1$) reveals the most dominant mode for transfer over the network.  This leading eigenvector is referred to as the {\it eigenvector centrality} $\boldsymbol{v}_1$.  The above eigenvector centrality is the out-eigenvector centrality $\boldsymbol{v}_1^\text{out}$.  It is also possible to determine the in-eigenvector centrality $\boldsymbol{v}_1^\text{in}$ by solving the eigenvalue problem $\boldsymbol{A}^T \boldsymbol{v} = \lambda \boldsymbol{v}$ for $\boldsymbol{A}^T$.

The eigenvector centrality for the undirected network example of Figure \ref{fig:example} is shown in Figure \ref{fig:strength}.  Compared to the localized degree centrality presented previously in Figure \ref{fig:example}, the eigenvector centrality uncovers the primary modal profiles of influential nodes.  This is highlighted by the capturing of the black node in the middle of the network acting as the dominant influence.  The dark red nodes on the right side of the network in fact can be captured as subdominant eigenvectors (not shown).

In addition to the eigenvector of the adjacency matrix, it is also useful to consider the leading eigenvector of the graph Laplacian $\boldsymbol{L}$.  Since the graph Laplacian is a diffusion operator, the leading eigenvector is sharper in its profile.  This is evident from the graph Laplacian eigenvector for the example network in Figure \ref{fig:strength}.  The graph Laplacian eigenvector accentuates the most influential nodes in the network.  

These eigenvector centralities are important for revealing the characteristics of networked dynamics and constructing a dynamical model.  As evident from equation (\ref{eq:Ax}), the eigenvector centrality has a close relationship with the linear stability analysis, discussed later in Section \ref{sec:linear_dynamics}.

\subsubsection*{Katz centrality and broadcast analysis}
\label{sec:Katz}

Assessing the influence of external forcing over a network is critical in many cases.  In network science, the {\it Katz centrality} \citep{Katz:Psychometrika53} identifies the influential nodes in such a setting.  Let us consider the evolution of a state variable $\boldsymbol{x}$ over a network with a forcing input $\boldsymbol{u}$
\begin{equation}
	\boldsymbol{x}^{m+1} 
	= \alpha \boldsymbol{A} \boldsymbol{x}^{m} 
	+ \boldsymbol{B} \boldsymbol{u}^{m},
\end{equation}
where $B$ represents the location over which input $u$ is introduced and $\alpha>0$ is a temporal parameter (down-weighting parameter).  If we consider the stable asymptotic behavior of the system, we can drop the temporal superscript and say
\begin{equation}
	\boldsymbol{x} = [ \boldsymbol{I} - \alpha \boldsymbol{A}]^{-1} \boldsymbol{B} \boldsymbol{u},
\end{equation}
which reveals the relationship between the input $\boldsymbol{u}$ and the state variable $\boldsymbol{x}$.  In network science, the Katz centrality $\boldsymbol{v}_\text{Katz}$ is usually determined by setting $B \boldsymbol{u} = \boldsymbol{1} = (1, 1, \dots, 1)^T$:
\begin{equation}
	\boldsymbol{v}_\text{Katz} = [ \boldsymbol{I} - \alpha \boldsymbol{A}]^{-1} \boldsymbol{1}.
\end{equation}
This of course can be generalized by considering an appropriate $B \boldsymbol{u}$ that is not constant over the network.  Here, we note that parameter $\alpha$ is chosen to be less than $1/\lambda_1$ to ensure stability \citep{Newman18}.  An example of the Katz centrality is shown in Figure \ref{fig:strength}.  In this particular case, we observe that the nodes in the middle and right two regions of the network are highly influential with forcing, similar to the degree distribution profile.

We can also consider extracting the leading forcing and response modes from the singular value decomposition of
\begin{equation}
	[ \boldsymbol{I} - \alpha \boldsymbol{A}]^{-1} 
	= \boldsymbol{U}_r \boldsymbol{\Sigma} \boldsymbol{V}^*_b,
\end{equation}
where $\boldsymbol{U}_r$ and $\boldsymbol{V}_b$ hold the receiving and broadcasting modes, respectively, over the network $\boldsymbol{A}$ and $\boldsymbol{\Sigma}$ contains the gains (singular values) in descending order \citep{Yeh:JFM21}.  This analysis is referred to as the {\it broadcast analysis}.  The leading pair of broadcasting and receiving modes $\boldsymbol{v}_{b1}$ and $\boldsymbol{u}_{r1}$, respectively, identify the most influential and influenced nodes.  

The Katz centrality and broadcast analysis have close connection to the transfer function in state-space representation of linear dynamics 
\begin{equation}
	\dot{\boldsymbol{x}} 
	= \boldsymbol{A} \boldsymbol{x} + \boldsymbol{B} \boldsymbol{u},
\end{equation}
which can be studied in frequency space with Laplace transform to yield
\begin{equation}
	\boldsymbol{x}(s) 
	= [s \boldsymbol{I} - \boldsymbol{A}]^{-1} \boldsymbol{B} \boldsymbol{u}(s),
	\label{eq:tf}
\end{equation}
where $s = \sigma + i \omega \in \mathbb{C}$.  Here, we have the transfer function (resolvent) $\boldsymbol{H} = [s \boldsymbol{I} - \boldsymbol{A}]^{-1} B$ that relates the input $\boldsymbol{u}(s)$ to the state variable (output) $\boldsymbol{x}(s)$.  The singular value decomposition of the resolvent operator $\boldsymbol{H}$ leads to the {\it resolvent analysis}, which is widely used in fluid dynamics \citep{Trefethen:Science93, Jovanovic:JFM05}.  Note that these is a close similarity between broadcast analysis and resolvent analysis in the determination of the modal profiles of the most influential and influenced structures.  The connections among resolvent analysis, broadcast analysis, and Katz centrality are discussed in further detail in \citep{Grindrod:PRSLA14, Aprahamian:JCN16, Yeh:JFM21} with extensions to analyze time-varying networks.

\subsection{Community detection}
\label{subsec:comdet}

As discussed above, centrality measures assess the relative influence of a single node in a network. However, for large networks, it is useful to discover groups or communities of nodes within the networks. Community detection is one of the most powerful tools in network science that groups the nodes with strong ties into a community.  The nodes within an identified community are tightly knit with as few edges as possible between different communities. 

If the number of characteristic communities in a network is known, then this leads to a graph partitioning problem. Local search strategies that rely on an arbitrary initial partition (e.g.,~Kernighan–Lin algorithm \citep{kernighan1970efficient}) and global strategies that rely on overall network properties (e.g.,~spectral partitioning \citep{mcsherry2001spectral}) can be employed for graph partitioning. Spectral partitions are derived from the approximate eigenvectors of the adjacency matrix. This type of approach shares similarities with grid partitioning that is commonly employed for parallel simulations of fluid flows to achieve minimal interprocessor communications.

In many cases, the number of communities in a network may not be known {\it a priori}. To search for naturally occurring groups in networked systems, one popular strategy is to maximize the modularity of a network \citep{newman2004fast}. The modularity of a network is defined as 
\begin{equation}
Q = \frac{1}{2m}\sum_{i,j} \left(A_{ij} - \frac{s_i^\text{in} s_j^\text{out}}{m}\right)\delta(c_i,c_j),
   \label{eq:modularity}
\end{equation}
where $m$ is the number of edges in the network, $c_i$ is the community index associated with node $i$ and $\delta$ is the Kronecker delta function. 

An exhaustive search of all possible groups increasing the modularity score $Q$ can be performed. However, this is computationally expensive and often intractable for large networks. Greedy optimization techniques alleviate these issues by iteratively optimizing the local communities until the modularity score can no longer be improved \citep{blondel2008fast}. In doing so, the community detection algorithms search for the optimal number of communities that maximize the modularity score of the network. 

\begin{figure}
	\centering
	\includegraphics[width=0.43\textwidth]{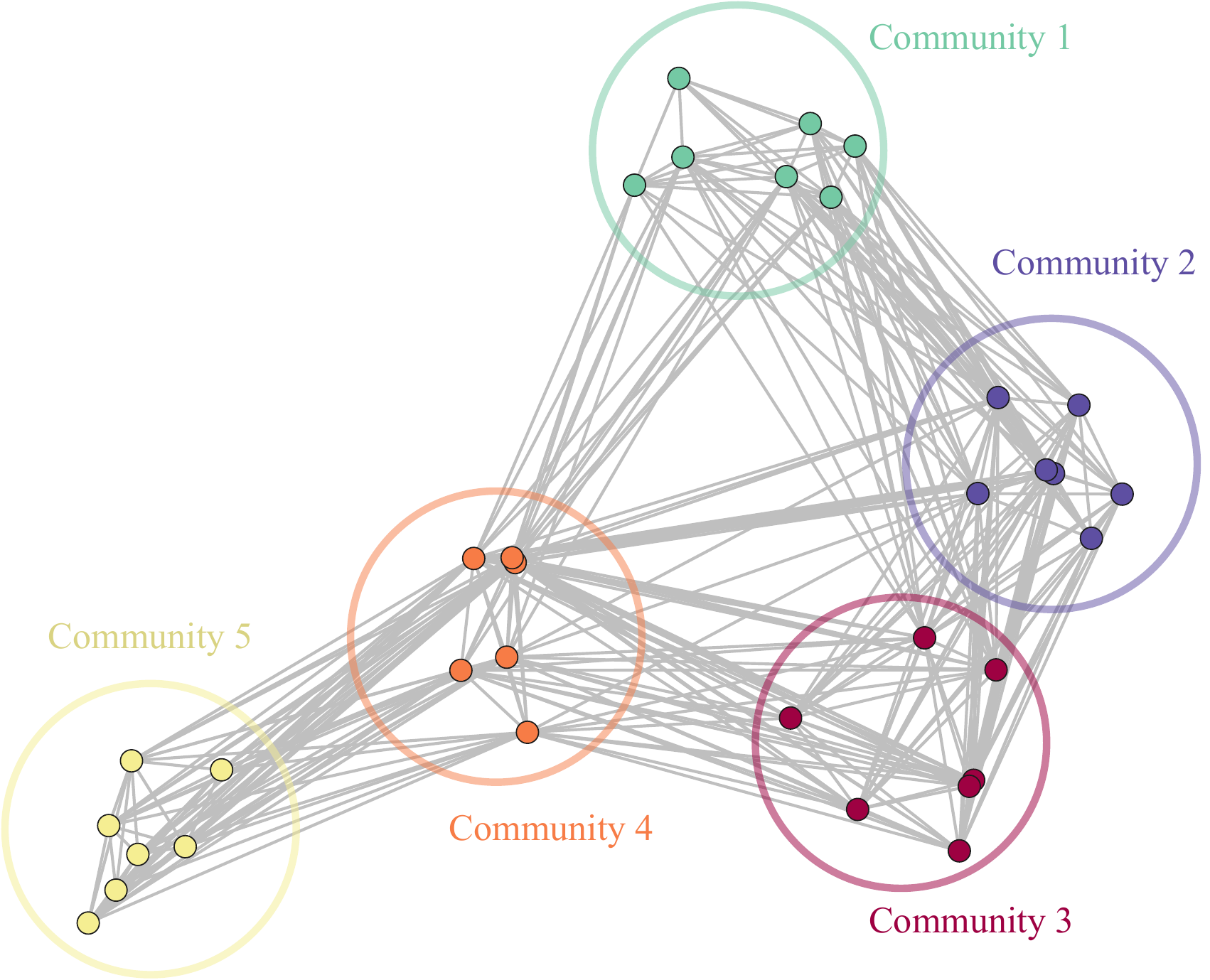}
	\caption{Communities detected from the example network through the modularity maximization algorithm.  Five communities are identified.
	}
	\label{fig:community}
\end{figure}

For the undirected network example of Figure \ref{fig:example}, the modularity maximization algorithm identifies five distinct communities as shown in Figure \ref{fig:community}. This community structure is also seen in the visualization of the adjacency matrix in Figure \ref{fig:example} (right) with stronger ties within each community and weaker ones between the communities. Besides modularity maximization, other methods including statistical inference \citep{holland1983stochastic} and clique-based methods \citep{palla2005uncovering} can also find network communities.

\subsection{Network models}

There is a diverse set of networks that represent interconnections studied in physical and social sciences.  We present here some of the key network models that possess important properties.  For simplicity, the examples herein are presented as unweighted and undirected networks but the discussions can be generalized to weighted and directed networks. For detailed discussions on network models, the readers should refer to \citep{Newman18, Barabasi16, Estrada12, Newman:SIAMReview03}.

\subsubsection{Basic networks}

Here, we present fundamental networks that serve as the backbone of basic network structures \citep{Newman18,Barabasi16,Estrada12}.

\begin{figure*}
	\centering
	\includegraphics[width=0.9\textwidth]{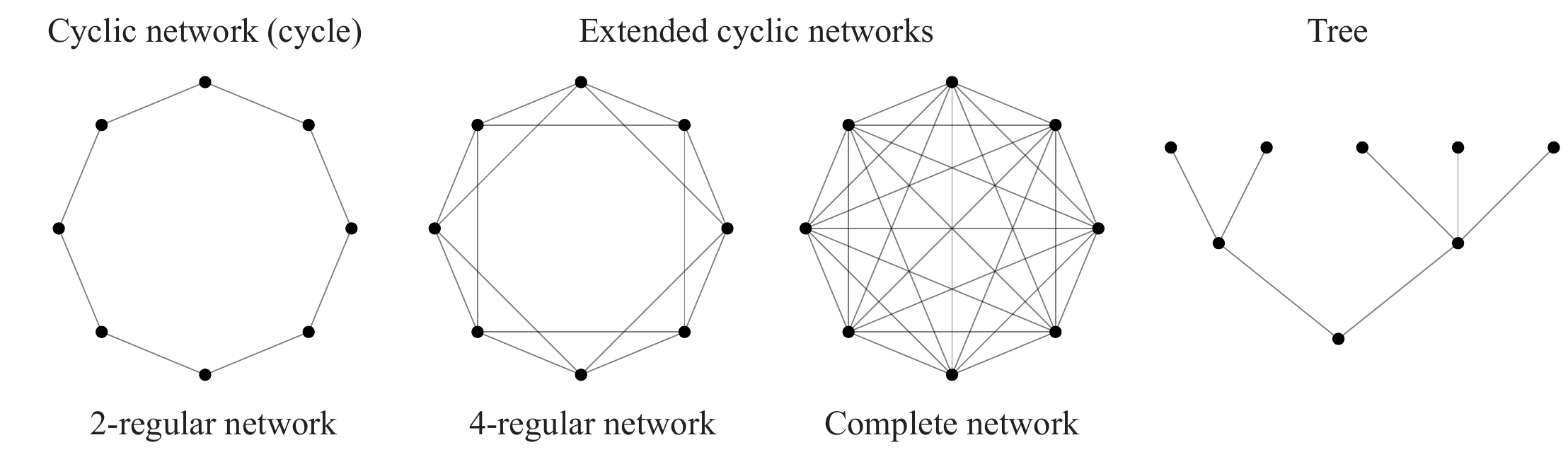}
	\caption{Basic graph examples.}
	\label{fig:basicgraphs}
\end{figure*}

\subsubsection*{Cyclic network}

If the connection among nodes on a network form a closed path, as shown in Figure \ref{fig:basicgraphs}, such a network is called a {\it cyclic network} or {\it cycle}.  Cyclic networks whose nodes with degrees higher than 2 are referred to as an extended cyclic network.  Examples of extended cyclic networks are also shown in Figure \ref{fig:basicgraphs}.

\subsubsection*{Regular and complete networks}

A {\it regular network} is comprised of a set of nodes, on which each node has the same number of edges.  If every node on a network has $k$ connections (degree $k$), it is called a $k$-regular network.  If each of the nodes on the network is connected to all other nodes, then such a network is referred to as a {\it complete graph}.  We show examples of regular and complete networks in Figure \ref{fig:basicgraphs}.

While the regular and complete network examples shown in the Figure are cyclic, regular networks, in general, do not need to be cyclic.  In fact, Cartesian grids are also regular networks.  If vertices on grid cells are considered to be network nodes, two and three-dimensional Cartesian grids would constitute $4$ and $6$-regular networks, respectively, excluding the boundary nodes \citep{Pozrikidis14}.

\subsubsection*{Tree}

A tree is a network that contains no closed path as shown in Figure \ref{fig:basicgraphs}.  This type of network appears often in databases.

\subsubsection{Random networks}

Let us also present three random network models that appear often in network science.  The three network models we describe below have connections between nodes based on particular probability distributions.  
These networks have structural properties which can then be used to describe networks encountered in various fields of engineering and science. 
In particular, we discuss here (i) the small-world network, (ii) the Erd\H{o}s-R\'{e}nyi network, and (iii) the scale-free network.

\subsubsection*{Small-world network}

In 1967, Stanley Milgram performed a famous experiment that asked randomly chosen people in Omaha, Nebraska, and Wichita, Kansas to send a letter to a particular individual in Boston, Massachusetts with one rule: The letter must be sent to a person that the sender knew on a first-name basis.  Through a chain of correspondence, the letter reached its final destination in Boston to determine how many times a letter needs to be forwarded to travel between two seemingly random persons.  Although the two originating cities in the Midwestern United States are physically far from Boston, it was found that only 5.5 or 6 steps of letter forwardings (average path length) were needed for the letters to reach their final destination \citep{Milgram:PT67,Travers:Sociometry69}. 

Most people only become acquainted with those in their neighborhoods (in the pre-internet age).  
However, once a person moves to a distant location, the social network starts to possess different characteristics. 
In fact, the transition from knowing only those in the local area to being able to pass on a letter over a long distance suggests that the social distance between any two people over a social network becomes very short.  
This concept was elegantly demonstrated through what is now known as the {\it small-world network} by Watts and Strogatz \citep{Watts:Nature98}.  

\begin{figure*}
	\centering
	\includegraphics[width=0.98\textwidth]{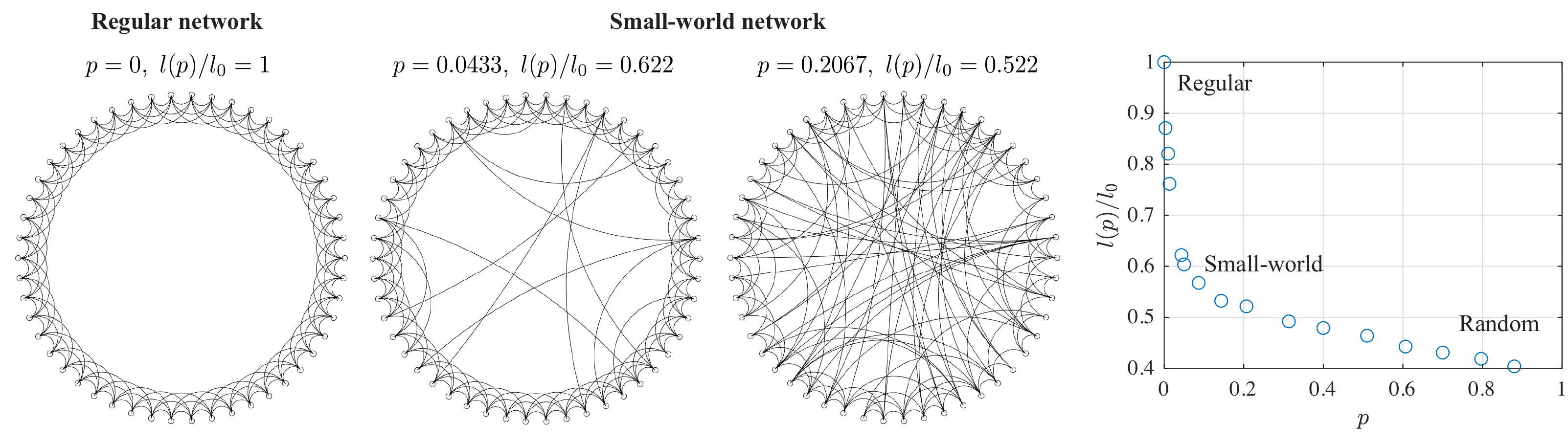}
	\caption{Generation of small-world network through rewiring a regular network.  Even with a small fraction $p$ of rewiring connections, the average distance $l(p)$ between pairs of nodes shows drastic reduction (normalized by mean distance of the regular network).}
	\label{fig:smallworld}
\end{figure*}

In this network model proposed by Watts and Strogatz, a regular network is rewired based on a prescribed probability.  Noteworthy here is that even with a low probability of rewiring of a regular network, the average path length (i.e., the number of transfers for the letter) can be dramatically reduced\footnote{This easily explains why COVID-19 has spread over the world so quickly with some of the population traveling globally.}.  An example of such drastic transition is shown in Figure \ref{fig:smallworld}.  For this given example of an extended cyclic graph, the average path length between vertices undergoes a significant reduction for a rewiring probability around $p = 0.04$.  This simple model describes the closeness of two vertices over a small-world network and is regarded as a cornerstone finding in network science.

\subsubsection*{Erd\H{o}s-R\'{e}nyi network}

\begin{figure*}
	\centering
	\includegraphics[width=0.98\textwidth]{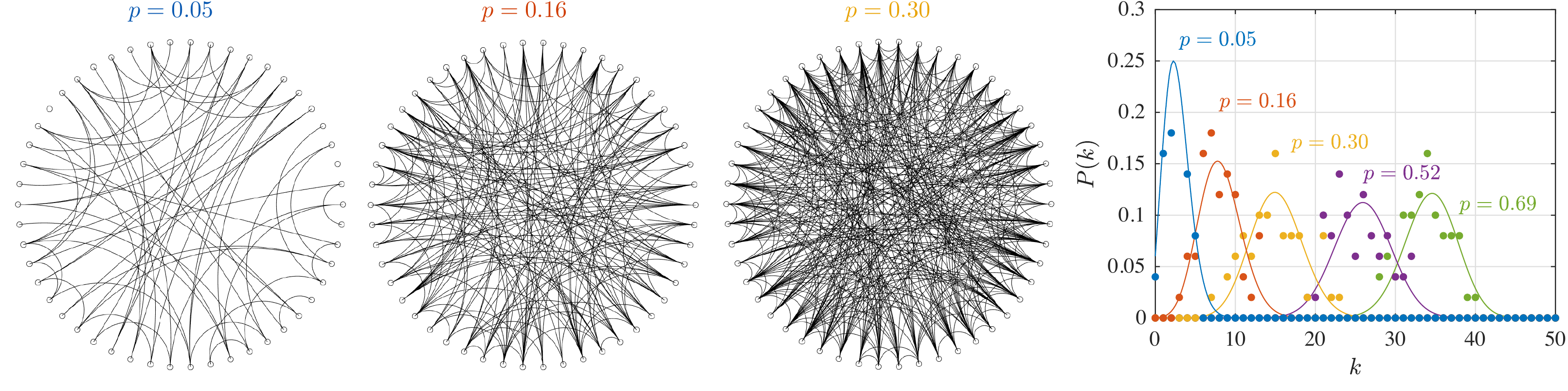}
	\caption{Erd\H{o}s-R\'{e}nyi networks visualized for different values of $p$.  Corresponding degree distributions for networks with $n=50$ are shown on the right with binomial distribution reference curves.}
	\label{fig:ERnetwork}
\end{figure*}

One of the most fundamental and classic (undirected) random networks is the {\it Erd\H{o}s-R\'{e}nyi model} \citep{ErdosRenyi1959, ErdosRenyi1960}.  This network model can be constructed by prescribing the probability $p$ for each pair of nodes in a network to be connected.  This network has a mean nodal degree of $\overline{k} = (n-1)p$ and a binomial distribution for its strength distribution
\begin{equation}
	P(k) = \begin{pmatrix} n-1 \\ k \end{pmatrix} p^k (1-p)^{n-1-k},
\end{equation} 
which approaches a Poisson distribution 
\begin{equation}
	P(k) = \frac{1}{k!} e^{-\overline{k}} \overline{k}^k
\end{equation}
for a large network ($n \gg 1$).  The Erd\H{o}s-R\'{e}nyi model is often referred to as the {\it classic random network} model.
An example of the Erd\H{o}s-R\'{e}nyi model is shown in Figure \ref{fig:ERnetwork}. Here, the classic random networks are generated for different values of $p$ for a fixed $n$.  The corresponding degree distributions are centered around their peaks at $\overline{k}$.

An interesting property of the Erd\H{o}s-R\'{e}nyi network is its ability to generate a large network component even with a very low probability $p$ between its nodes.  This can be seen from Figure \ref{fig:ERnetwork}, where almost all nodes are connected to the network.  In fact, it is known that there is a dramatic transition from the majority of the nodes being disconnected to almost all the nodes being well-connected over the network when $pn \approx 1$ \citep{ErdosRenyi1960}.  This property tells how easily a network can extend to the majority of the population as long as there is some probability for an individual node to be connected to another node.

This type of model describes physical networks such as the highway system.  Cities are connected to nearby cities and generally have a similar number of connections, which amounts to having a Poisson-like distribution about the mean number of connections or highways \citep{Barabasi:SA03}.

\subsubsection*{Scale-free network}

The Erd\H{o}s-R\'{e}nyi network model assumes that the connection between nodes has a uniform probability $p$.  However, such an assumption is not valid for many networks.  Here, we consider the probability of having a connection between a pair of nodes to be a function of the degree distribution.  For this network, we start with an initial network to which we add new nodes according to $p = k/\sum_i k_i$.  This means that new nodes will have tendencies to be connected to well-connected nodes.  This model is known as the Barab\'{a}si--Albert model \citep{BarabasiScience99} and essentially amounts to the rich gets richer phenomena (preferential attachment model).  The network generated by this model results in a power-law degree distribution
\begin{equation}
    \mathcal{P}(k) = k^{-\gamma},
\end{equation}
which is known as the {\it scale-free network}.  Here, $\gamma$ is the exponent of the scale-free distribution, which generally takes $2 < \gamma < 3$ for many networks \citep{Newman:SIAMReview03, Barabasi16, Newman18, Taira2016jfm}.  The discussion here is based on a growing network but there exist other models that generate the scale-free network through a stationary perspective \citep{Newman:PRE01}.  

\begin{figure*}
	\centering
	\includegraphics[width=0.98\textwidth]{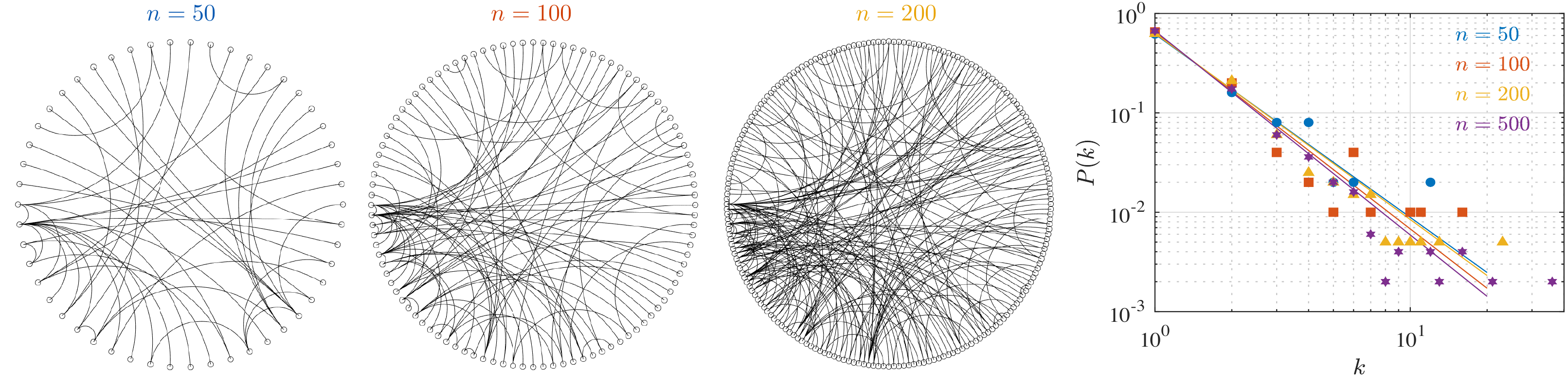}
	\caption{Scale-free networks visualized for increasing number of nodes.  Corresponding degree distributions for networks with $n=50$, $100$, $200$, and $500$ are shown on the right with power-law reference curves.}
	\label{fig:SFnetwork}
\end{figure*}

We present an example of the scale-free network and its degree distribution in Figure \ref{fig:SFnetwork}.  As the name suggests, there is no characteristic degree for a scale-free network, in contrast to the peaky behavior exhibited by the Erd\"{o}s-R\'{e}nyi network.  Instead, the scale-free network possesses a fat-tail distribution due to the small number of nodes with high nodal degrees on the network.  These highly-connected nodes with high degrees are known as {\it hubs} and can be seen in the figures with a large number of connections (see the bottom left side of the networks).  For the shown example, the power-law distributions have $\gamma \approx 2$.  While we show the examples in Figure \ref{fig:SFnetwork} for up to $n = 500$, there are some fluctuations in the presented power-law behavior.  These fluctuations will be lower for larger networks and with the use of larger binning sizes for evaluating the degree distributions.   

Scale-free networks are encountered in a wide range of networks, including the airline networks, the world wide web, and social networks \citep{Newman18,Estrada12}.  The airline networks for major US airlines illustrate the scale-free characteristics.  There are only a handful of hub airports that are highly connected to many airports.  However, there are a large number of medium to small-sized airports across the US that are mainly connected to the hubs.  These characteristics are different from the US highway network that we discussed for the Erd\"{o}s-R\'{e}nyi network \citep{Barabasi:SA03}.  Moreover, for the airline networks, new routes are usually serviced from the hub airports following the preferential attachment perspective.

The scale-free network is known to be resilient against random perturbations but not for targeted attacks against the hubs.  This point has important implications for control of dynamics on networks \citep{Barabasi:SA03,Albert:Nature00,Taira2016jfm}.  This property explains why the airline networks operate well despite the occurence of minor perturbations throughout the network.  However, airline operations can fail catastrophically when adverse weather hits hub airports.

\subsection{Networked dynamics}

The evolution of interactions can be described as dynamics on a network.  
Decomposing the dynamics into intrinsic and interactive components, we can express the governing equation for variable ${\varphi}_i$ as 
\begin{equation}
   \dot{{\varphi}}_i 
   = {f}({\varphi}_i) 
   + \sum_{j=1}^n A_{ij} \boldsymbol{g}({\varphi}_i, {\varphi}_j),
   \quad i = 1, 2, \dots, n 
   \label{eq:networkdyn}
\end{equation}
where ${f}$ and ${g}$ are nonlinear functions and $A_{ij}$ is the adjacency matrix \citep{Newman18}.  Function ${f}({\varphi}_i)$ represents the intrinsic dynamics at node $i$ and function ${g}({\varphi}_i,{\varphi}_j)$ describes the interactive dynamics between adjacent nodes $i$ and $j$. In general, these functions can also be dependent on $i$ and $j$, such as ${f}_i({\varphi}_i)$ and ${g}_{ij}({\varphi}_{ij})$.  
In the case of the diffusive process given by Eq.~(\ref{eq:diffusion}), ${g}({\varphi}_i, {\varphi}_j) = \alpha({\varphi}_i - {\varphi}_j)$. 
Apart from diffusion processes, other discrete and continuous-state dynamical processes have been studied on networks. These include networked stochastic processes \citep{holland1983stochastic, wasserman1980analyzing}, synchronization processes \citep{arenas2006synchronization}, coordination games \citep{jackson2015games, mccubbins2009connected}, and fluid flow interactions \citep{Nair:JFM15, meena2018network, nair2018network}.

To further elucidate the role of network structure and dynamics on the evolution of the variables, let us assess the stability of the networked dynamical systems about the equilibrium state $\varphi_i^*$.  We consider the state $\varphi_i = \varphi_i^* + \epsilon_i$ to be perturbed with small perturbation $\epsilon_i$ about the equilibrium.  Linearizing  Eq.~(\ref{eq:networkdyn}) about $\varphi_i^*$, we obtain 
\begin{equation}
\begin{split}
\frac{\mathrm{d}{\epsilon}_i}{\mathrm{d}t} 
= \epsilon_i \frac{\partial f}{\partial \varphi}\Bigr|_{\varphi = \varphi_i^*} 
& + \epsilon_i \sum_j A_{ij}\frac{\partial g(u,v)}{\partial u}\Bigr|_{\substack{u = \varphi_i^* \\ v = \varphi_j^*}} \\
+ & \sum_j A_{ij}\epsilon_j\frac{\partial g(u,v)}{\partial v}\Bigr|_{\substack{u = \varphi_i^* \\ v = \varphi_j^*}} 
   \label{eq:networkdyn11}
\end{split}
\end{equation}
where the higher-order terms have been dropped from the Taylor expansion \citep{Newman18}. The right-hand side of the above expression can be rearranged in the matrix form to determine the stability of the fixed points. 

In the case of a diffusive process where $g({\varphi}_i, {\varphi}_j) = g({\varphi}_j) - g({\varphi}_i)$ with a symmetric equilibrium $\varphi_i^* = \varphi^* $, the analysis can be reduced \citep{arenas2006synchronization} to 
\begin{equation}
\frac{\mathrm{d}{\epsilon}_i}{\mathrm{d}t} = \mathcal{J}[f(\varphi^*)] \epsilon_i -  \mathcal{J}[g(\varphi^*)]\sum_j L_{ij}\epsilon_j,
   \label{eq:networkdyn12}
\end{equation}
where $\mathcal{J}[\cdot]$ is the Jacobian matrix. The above expression can be diagonalized to $\xi_l = (\mathcal{J}[f(\varphi^*)] - \lambda_l\mathcal{J}[g(\varphi^*)])\xi_l$ where $\xi_l$ is the eigenmode associated with the eigenvalue $\lambda_l$ of the Laplacian matrix $L_{ij}$. In general, the associated eigenvalues for weighted directed graphs are complex. For undirected  graphs associated with only real eigenvalues, the maximum eigenvalue of the function $\bs{\sigma}(\lambda_r) = \mathcal{J}[f(\varphi^*)] - \lambda_r\mathcal{J}[g(\varphi^*)])$ is called as the master stability function. The system is stable if all eigenvalues of this function are non-positive. The master stability function has the ability to decouple the role of structure and dynamics of networked dynamical systems \citep{pecora1998master}.

\subsection{Modeling networked dynamics}

It is often desirable to reduce the complexity of a networked dynamical system for modeling and extracting its essential dynamics on a large network.
Here, we discuss two approaches to derive networked dynamics models: (i) reduced-order models based on network communities \citep{meena2018network} and (ii) sparse network models based on edge removal \citep{Nair:JFM15}.  These two approaches are founded on reducing the complexity of a given network based on its structural connectivity and can be combined with other model reduction techniques for fluid flows \citep{Holmes96, Antoulas05, Noack11, Rowley:ARFM17, Taira_etal:AIAAJ17, Taira_etal:AIAAJ20, nair2018network, Mou:Fluids21}.  These models can also serve as the basis to perform feedback control of the networked dynamics.

\subsubsection{Community-based reduction}

Tracking the dynamics at every node on a large network is often intractable or unnecessary.  In such a case, we can model the dominant dynamics that take place amongst network communities at a macroscopic level.  This approach reduces the state variable dimension of the networked dynamical system by grouping the nodes of the network into communities, identified by community detection from \S\ref{subsec:comdet} or prior knowledge. 

To consolidate individual dynamics within a community, we define a community centroid variable $\tilde{{\varphi}}_k$ as a weighted average of the nodes within a community $c_k$, 
\begin{equation}
    \tilde{{\varphi}}_k 
    \equiv 
    \frac{\sum_{i\in c_k}\kappa_i{\varphi}_i}{\sum_{i\in c_k}\kappa_i}, 
    \label{eq:centroid}
\end{equation}
where $\kappa_i$ is a variable of interest that contributes of the individual nodes to the community centroid (e.g., vorticity can be used as $\kappa_i$ in vortex-dominated flows).  
The reduced networked dynamics model of the community centroids is obtained by performing a community-based average of Eq.~(\ref{eq:networkdyn}) to yield
\begin{equation}
    \dot{\tilde{{\varphi}}}_i = \tilde{{f}}(\tilde{{\varphi}}_i) + \sum_{j=1}^m \tilde{A}_{ij} 
    \tilde{{g}}(\tilde{{\varphi}}_i, \tilde{{\varphi}}_j),
    \quad i = 1, 2, \dots, m \ll n,
    \label{eq:centroid_dynamics}
\end{equation}
where $\tilde{{f}}$ and $\tilde{{g}}$ are the centroid-based nonlinear functions and $\tilde{\boldsymbol{A}}$ is the community-reduced adjacency matrix \citep{meena2018network}.  Because the number of communities $m$ is much smaller than the total number of nodes $n$, the reduced-order networked dynamics model is a low-rank approximation of the full dynamics.

\subsubsection{Network sparsification}

An alternate way to reduce the network description is to perform sparsification of the underlying graph.  
For dense networks, edges can be removed while keeping important connections.  While removing edges, the resulting sparse networks should maintain similar properties as the original network. We note that simple removal of edges based on thresholding (i.e., removing edges with $A_{ij}$ below some threshold) is generally not recommended since the sparsified network likely will not conserve important properties of the networked dynamics \citep{Nair:JFM15}.  In many cases, the removed edge weights should be redistributed to other edges that remain in the sparse representation.  

There are different sparsification strategies based on different notions of graph similarity.  Reducing the number of edges while keeping the same shortest-path distance between nodes leads to the notion of distance similarity \citep{peleg1989optimal} and maintaining the out-degree of the nodes of the network leads to cut similarity \citep{benczur1996approximating}.  A more robust notion of similarity is that of spectral similarity \citep{spielman2011graph, spielman2011spectral}. In this approach, spectral sparsification can find sparse networks while maintaining the eigenspectra, which is critical for accurately modeling the dynamical systems.

Once we obtain a sparse network, we can approximate the dynamics of the networked system as
\begin{equation}
   \dot{{\varphi}}_i = {f}({\varphi}_i) 
   + \sum_{j=1}^n A^s_{ij} {g}({\varphi}_i, {\varphi}_j).
   \quad i = 1, 2, \dots, n. 
   \label{eq:sparse_networkdyn}
\end{equation}
Here, $\boldsymbol{A}^s$ is the adjacency matrix for the sparse network.  While this governing equation describes the dynamics for a state variable that has the same size as that for the original dynamics, the interactive dynamics take place over a lower number of edges.  When a large number of edges are removed, the sparse dynamics model achieves a significant reduction in the required computational effort as the summation in \eq (\ref{eq:sparse_networkdyn}) calls for a much fewer number of summands.  One can also combine this sparse network model with the community-based reduction model from \eq (\ref{eq:centroid_dynamics}) to derive a low-order sparse model for networked dynamical systems.  An example of a sparse network model for vortical flow is presented later in Section \ref{vorticalnetwork}.

\subsection{Network inference}

Network-based analyses rely on having access to network structures.  For interactions that are well-defined or those that can be derived from physics, the construction of a network is straightforward.  However, there are many cases in which an unknown network needs to be determined from measurements.  This constitutes a reconstruction problem known as network inference to estimate the network structure from observable data. 

Network inference requires the estimation of the network structure from data and optimizing the structural parameters to accurately reproduce networked dynamics \citep{hecker2009gene}. For estimating the network structure, statistical dependencies in the data including correlations \citep{opgen2007correlation}, causality \citep{granger1969investigating, friston2011functional} and other information theory metrics \citep{schreiber2000measuring} are commonly exploited.  In addition to detecting direct interactions, we should be aware that indirect interactions between the network nodes may be inferred due to the de-correlating effect of the other nodes, which poses challenges.  Recently, there are ongoing studies to establish a noise-robust inference of networks based on differential equations and Bayesian approaches \citep{nawrath2010distinguishing, casadiego2017model,aubin2020gene, saint2020network}. Bayesian networks provide a flexible framework by combining prior knowledge to handle noisy measurements and hidden variables. 

To reproduce the networked dynamics accurately, both explicit and implicit optimization techniques can be employed to determine the network structure \citep{hecker2009gene}. Explicit techniques optimize the network parameters based on explicit scoring functions, such as Bayesian information criteria \citep{schwarz1978estimating} and Akaike information criterion \citep{akaike1998information}, to improve network sparseness.  On the other hand, implicit optimization approaches penalize the number of interactions by regularization (e.g., ridge \citep{hoerl1970ridge} and LASSO \citep{tibshirani1996regression}) to find robust network connections. 
We note that there is often an over-reliance on the type, quality, and amount of the data with these methods which necessitates careful experimental designs for generating the data to be utilized. By perturbing the system dynamics and careful sampling strategies, the data quality and subsequently the accuracy of network inference can be significantly improved.


\section{Network analysis of fluid flows}
\label{sec:flows}

\begin{table*}
{
    \caption{An overview of representative network-based approaches for the analysis of fluid flows.}
	\label{table:summary}
	\centering
	\begin{tabular}{p{4cm}p{3.1cm}p{2.7cm}p{3.2cm}p{1.25cm}} 
	\hline 
	{\textbf{Network approach}}       	& {\textbf{Node}}           & {\textbf{Edge}}	   	    & {\textbf{Weight}} 			& {\textbf{Sections}} \\ 
	\hline 
	\hline 
	{\textbf{Flow field network}}		& 					        & 					        & 								& \ref{sec:flow_field_network} \\ 
	\hline 
    Vortical interaction network	    & Vortex                 	& Vortical interaction   	& Induced velocity  			& \ref{vorticalnetwork} \\
	Linear dynamic network			    & State variable element	& Linear interaction		& Linear dynamics				& \ref{sec:linear_dynamics} \\
    Modal interaction network           & Spatial mode            	& Modal interaction         & Energy transfer 				& \ref{sec:modal_interaction} \\
	Lagrangian transport network        & Lagrangian element     	& Proximity                 & Distance based  				& \ref{particleproximity} \\
    \hline 
    {\textbf{Sensor network}} 			& 				        	& 					        & 								& \ref{subsec:tsn} \\
    \hline 
	Visibility graph  					& Scalar state              & Visibility 		        & Unweighted visibility			& \ref{sec:visibility} \\
    Recurrence network                 	& Time-delay state 		    & State recurrence 		    & Frequency of recurrence       & \ref{sec:recurrence} \\
    Cluster transition network    		& Cluster  		            & Cluster transition 		& Transition probability       	& \ref{sec:cluster_network} \\
    \hline 
	\end{tabular}
	}
\end{table*}

Network science can offer a refreshing perspective to analyze the complex dynamics and coherent structures of fluid flows.  Since the mathematical description of networks is general in nature, various aspects of networks embedded within fluid flows can be analyzed to understand their connectivities, structures, and dynamics.  With flexibility in the choice of nodes, edges, and edge weights, there is tremendous potential to gain physical insights into the structures and dynamics of the complex flow physics that may have been challenging to analyze with traditional methods.  While the use of network science is fairly recent in the field of fluid dynamics, we are now seeing developments of network-inspired approaches to analyze a variety of fluid flows over the past few years.  

One of the major questions in using network science for fluid dynamics is how to apply network analysis techniques to fluid flow analysis.  Networks are generally described as discrete entities, while fluid flows are described as continua.  While this may be subtle, it is a critical distinction that must be realized.  The mathematical framework of network analysis is generally based on finite-dimensional graph theory.  On the other hand, the description of a fluid flow (continuum) or its governing partial differential equations must first be discretized in an appropriate manner such that network science can aid the analysis.  Fluid flows also tend to be dense in their interconnections while common networks (in network science) are usually sparse.  These distinctions lead to differences in network properties, numerical approaches, and necessary computational resources.  Nonetheless, recent advances in data-driven techniques and graph theory are further supporting the use of network science applications for the analysis of fluid flows.  

In this section, we introduce some of the recent applications of network-driven approaches to analyze, model, and control fluid flows.  These works offer stimulating insights and approaches that may be transferable to a range of fluid flow problems.  The breadth of these studies is offered by considering the interactions or connections among different types of flow entities on networks.  We mainly categorize the uses of networks in two ways, as summarized in table \ref{table:summary}.  The first approach (Section \ref{sec:flow_field_network}) encompasses physical elements of the flow field (e.g, vortices, particles, and modes) to be nodes on the network with spatial information playing an important role.  The second approach (Section \ref{subsec:tsn}) utilizes sensor measurements to establish networks without the spatial information necessarily involved.  The present classification is a way to systematically list the family of methods that have been developed to date and will likely evolve as the fluid mechanics community makes advancements in extending the use of network science.  In fact, some of these classifications are becoming blurred as data-driven techniques bridge across many techniques.  In what follows, we highlight some of the network-based approaches in fluid mechanics and discuss potential extensions in this emerging field of research.


\subsection{Flow field network}
\label{sec:flow_field_network}

Let us survey network-based analysis techniques that leverage spatial information of the flow field.  This type of approach derives networks based on the flow field, such as the vorticity field or the Lagrangian particle trajectories.  By embedding the spatial and temporal insights of the flow into the network, it can be used to develop a reduced-order and sparse dynamical model for a range of vortical flows.  In this subsection, we discuss the vortex interaction network, linear dynamic network, modal interaction network, and Lagrangian transport network.

\subsubsection{Vortex interaction network}
\label{vorticalnetwork}

We start the discussion of the network-based analysis of fluid flows with one of the most fundamental flow interactions, namely, vortex interactions.  As an illustrative example, let us consider a collection of potential vortices.  By viewing each potential vortex to be a node and the vortical interaction to be a network edge, vortical interactions constitute a dense weighted network without self-loops.  

On this vortex interaction network, the edge weights can be assigned with the magnitudes of the induced velocity that a vortex imposes on other vortices.  For example, let us consider the case of two-dimensional flow with a set of $n$ potential vortices, where each vortex holds vorticity  
\begin{equation}
	\omega_i(\boldsymbol{r}) = \gamma_i \delta(\boldsymbol{r} - \boldsymbol{r}_i),
	\quad
	i = 1, \dots, n,
\end{equation}
where $\boldsymbol{r}_i$ and $\gamma_i$ are the position and strength (circulation) of vortex $i$, respectively, and $\delta(\boldsymbol{r})$ denotes the Dirac delta function.  The motion of the vortices is governed by the Biot--Savart law for this setup
\begin{equation}
	\frac{d\boldsymbol{r}_i}{dt} 
	= \sum_{j=1,j\ne i}^n \frac{\gamma_i}{2\pi} 
	\frac{\hat{\boldsymbol{e}}_z \times
	(\boldsymbol{r}_i - \boldsymbol{r}_j)}{\| \boldsymbol{r}_i - \boldsymbol{r}_j \|_2^2},
	\label{eq:BSlaw}
\end{equation}
where we have assumed the absence of external velocity input.  Here, the unit vector $\hat{\boldsymbol{e}}_z$ is oriented out of the plane of interest. 

\begin{figure*}
\centering
	\includegraphics[width=0.9\textwidth]{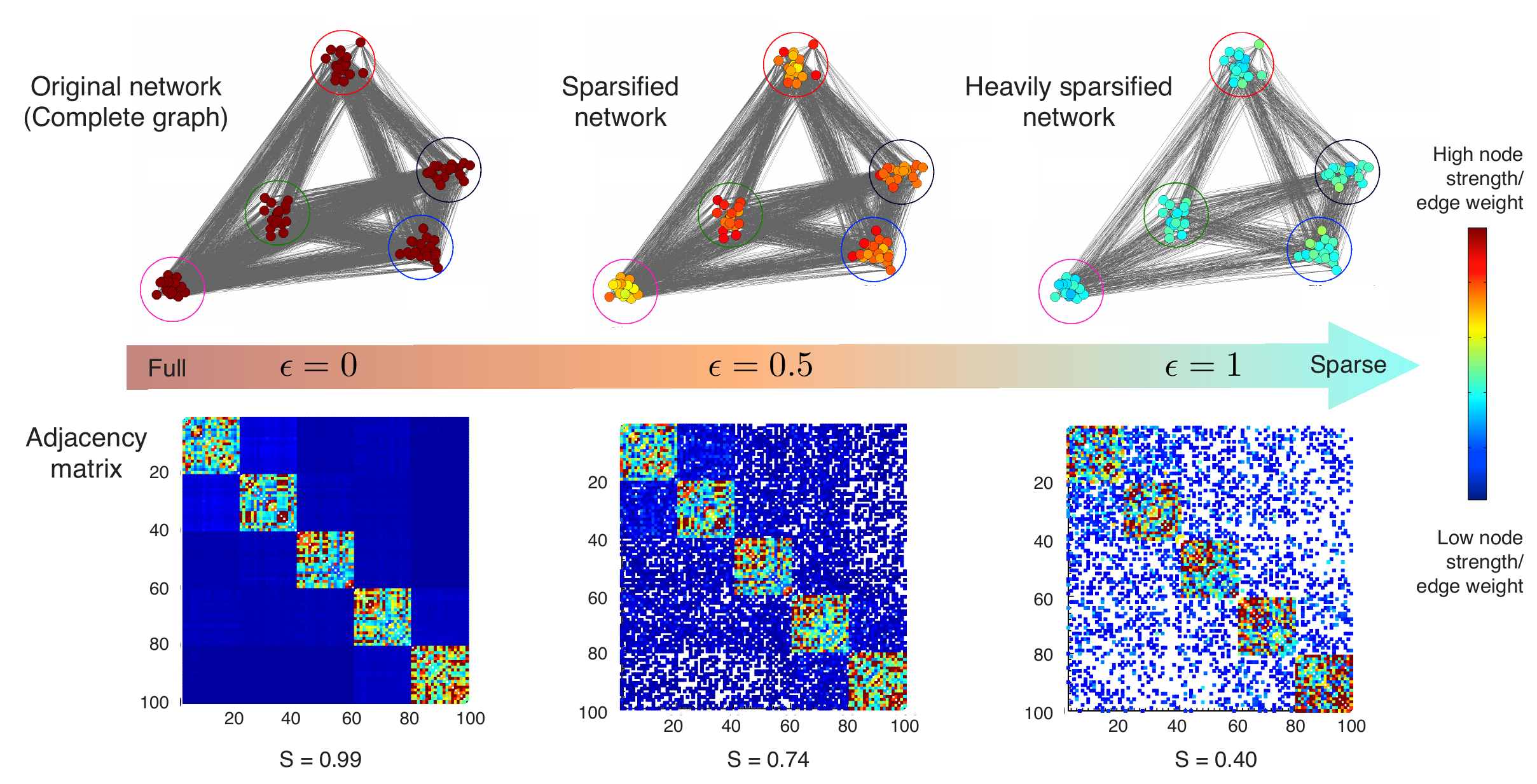}
	\caption{Vortical interaction network for $100$ potential vortices with random strengths and positions \citep{Nair:JFM15}.  Spectral graph sparsification results are shown for approximation order of $\epsilon = 0$ (original), $0.5$, and $1$ (heavily sparsified) yielding the shown sparsity factor $S$, which is the number of nonzero elements scaled by the total number of elements of the adjacency matrix.
	Reprinted with permission from Cambridge University Press.}
	\label{fig:vort_network}
\end{figure*}

Every vortex induces velocity on all other vortices, except the self-induced velocity on itself, as shown in Figure \ref{fig:vort_network}.
From equation (\ref{eq:BSlaw}), we observe that the magnitude of the induced velocity imposed by vortex $j$ onto vortex $i$ is 
\begin{equation}
	| u_{ij} | = 
	\begin{cases}
		\frac{\gamma_j}{2\pi \| r_i - r_j\|_2}, & i \ne j\\
		0, & i = j.
	\end{cases}
\end{equation}
Given the magnitude of influence in terms of the induced velocity, the edge weight can be accordingly defined using the arithmetic mean
\begin{equation}
	w_{ij} = \alpha | u_{ij} | + (1-\alpha) | u_{ji} | 
\end{equation}
or the geometric mean
\begin{equation}
	w_{ij} = | u_{ij} |^\alpha  | u_{ji} |^{(1-\alpha)} ,
\end{equation}
where parameter $0 \le \alpha\le 1$ is introduced for generality.  When $\alpha = 1$, we retrieve a directed network with $w_{ij} = | u_{ij} |$.  In the case of $\alpha = 1/2$, we obtain a symmetric network.  While a directed network is often appropriate to capture the physical interactions, the use of symmetric network is sometimes called for to apply network analysis techniques \citep{spielman2011graph}.  We note that this formulation can also be extended to three-dimensional vortical flow \citep{MGM:JFM21}.

As an example, let us consider a collection of $100$ potential vortices with random strengths and positions, as presented in Figure \ref{fig:vort_network} (top left).  The spatial arrangement of the vortices in this particular example is comprised of 5 clusters \citep{Nair:JFM15}.  
This is visible from the adjacency matrix shown in Figure \ref{fig:vort_network} (bottom left).  The 5 subblocks along the diagonal indicate stronger intra-cluster interactions.  Note that the adjacency matrix is dense but lacks any diagonal elements due to the absence of self-induced velocities.   
We can sparsify the vortex interaction networks to extract the key interactive networks as shown in Figure \ref{fig:vort_network} (middle and right).  
Such sparse representation has been used to model the overall dynamics of the vortices while conserving physical properties \citep{Nair:JFM15}.  
Community detection algorithms can also be utilized to determine a group of vortices that are arranged in a spatially cohesive manner. 
The extracted communities of vortices can be reduced to their centroids about which reduced and sparse representation of vortex dynamics can be derived \citep{Nair:JFM15,meena2018network}.

\begin{figure}[ht!]
\centering
	\includegraphics[width=0.48\textwidth]{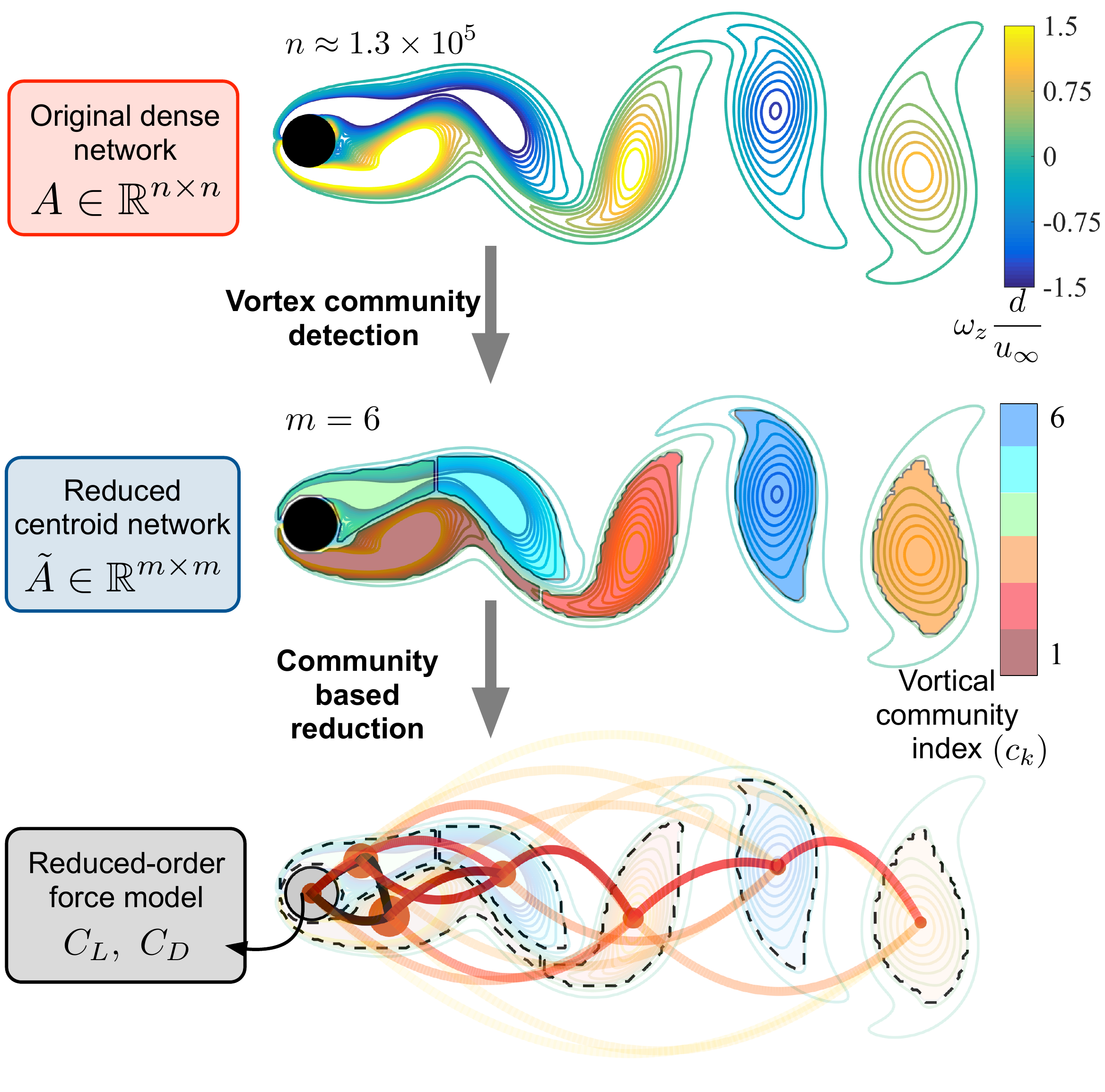}
	\caption{Network based modeling for cylinder wake vortices and forces \citep{meena2018network}.
	Reprinted with permission from the American Physical Society.
	}
	\label{fig:cylinder_comm}
\end{figure}

\begin{figure}[ht!]
\centering
	\includegraphics[width=0.42\textwidth]{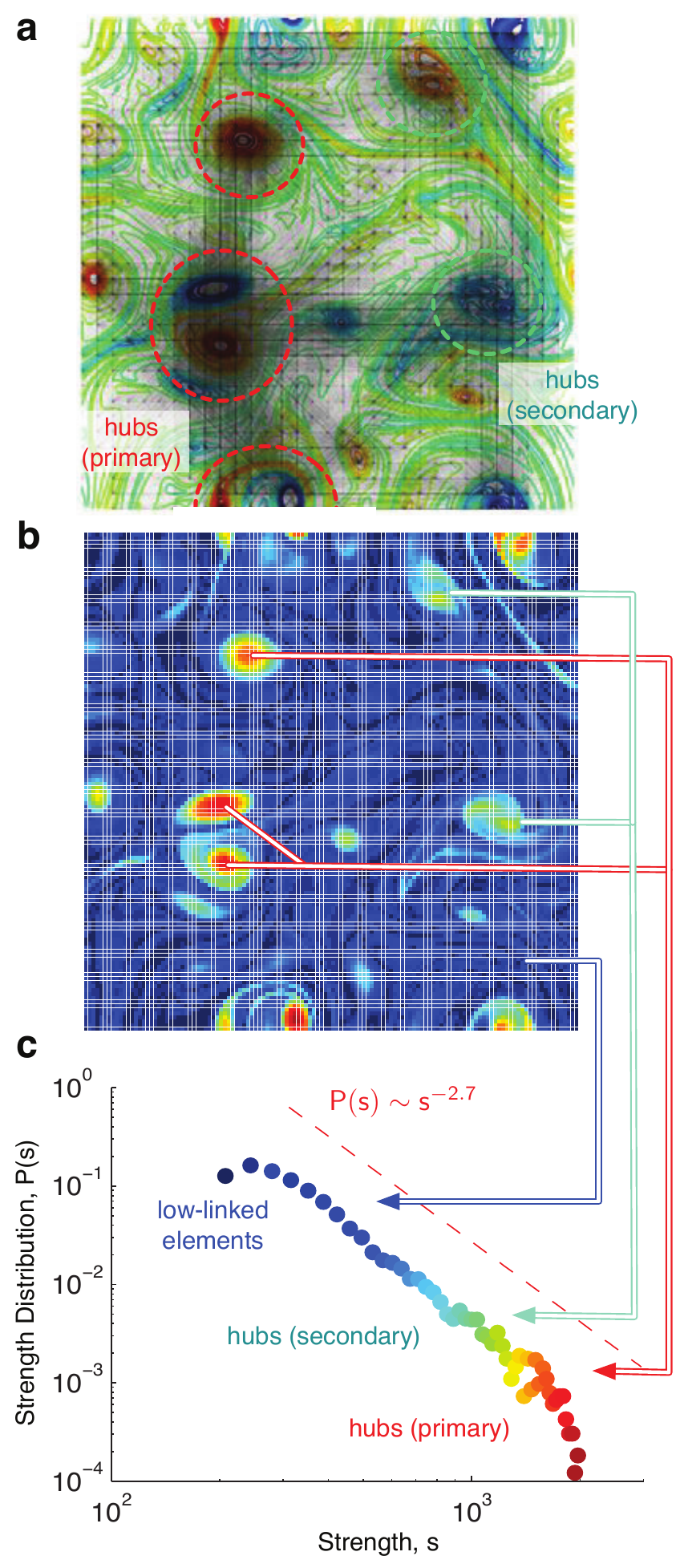}
	\caption{Vortical interaction network for two-dimensional isotropic turbulence \citep{Taira2016jfm}.  (a) Vortex interaction network, (b) node strength, and (c) scale-free strength distribution.  Reprinted with permission from Cambridge University Press.}
	\label{fig:2dturb}
\end{figure}

The above vortical network analysis is based on a Lagrangian perspective.  Alternatively, spatial grid points (fluid elements) can serve as network nodes through a Eulerian approach.  
In this case, the interactions between the vortical element $i$ on a grid cell and the vortical element $j$ on another grid cell are captured \citep{Taira2016jfm}.  By extending the vortex interaction formulation to a Eulerian discretization, numerical simulations and experimental measurements on Eulerian grids can be analyzed and be used for developing reduced-order models as shown in Figure \ref{fig:cylinder_comm} \citep{meena2018network}.  
This type of analysis has been applied to bluff body wakes \citep{meena2018network} and isotropic turbulence \citep{Taira2016jfm,MGM:JFM21}.  
Interestingly, it was shown that two and three-dimensional isotropic turbulence respectively exhibit scale-free and log-normal network characteristics.  The vortical network and its scale-free strength distribution for two-dimensional isotropic turbulence are shown in Figure \ref{fig:2dturb}.  
Such observations are insightful in identifying influential vortical structures and assessing the robustness of the flow again added perturbations \citep{Taira2016jfm}.  
Vortex interactions analysis has also been applied to reacting flows to identify sensitive regions that can modify the combustion dynamics \citep{murayama2019attenuation}.

\subsubsection{Linear dynamic network}
\label{sec:linear_dynamics}

There is a close connection between the descriptions of dynamical systems and networks.  Here, we provide perspectives on how network-based analysis techniques can be related to linear analysis techniques, including stability and resolvent analyses.  The connections will be discussed in terms of the formulation and the centrality measures discussed in \S\ref{centralitymeasures}. 

Let us first start by considering a fluid flow described by
\begin{equation}
	\frac{\partial \boldsymbol{q}}{\partial t} 
	= \boldsymbol{F}(\boldsymbol{q}) + \boldsymbol{f},
	\label{eq:}
\end{equation}
where $\boldsymbol{q}$, $\boldsymbol{F}$, and $\boldsymbol{f}$ represent the flow state, nonlinear evolution operator, and external forcing, respectively.  The flow state variable $q$ can be decomposed as
\begin{equation}
	\boldsymbol{q}(\boldsymbol{x},t) 
	= \overline{\boldsymbol{q}}(\boldsymbol{x},t) 
	+ \boldsymbol{q}'(\boldsymbol{x},t),
	\label{eq:decomp}
\end{equation}
where $\overline{\boldsymbol{q}}(\boldsymbol{x},t)$ is some base flow and $\boldsymbol{q}'(\boldsymbol{x},t)$ is the perturbation about $\overline{\boldsymbol{q}}$.  If the base flow is taken to be a time-invariant equilibrium flow and linearize the dynamics about the base state, we have
\begin{equation}
	\frac{\partial \boldsymbol{q}'}{\partial t} 
	= L_{\overline{\boldsymbol{q}}} \boldsymbol{q}' + \boldsymbol{f}',
	\label{eq:dyn_f}
\end{equation}
where we have assumed the magnitude of $\boldsymbol{q}'$ to be small and neglected nonlinear terms.  Here, the operator $\boldsymbol{L}_{\overline{\boldsymbol{q}}}$ represents the linearized operator.  If we are interested in the manner in which flow elements interacts linearly, we can take $\boldsymbol{L}_{\overline{\boldsymbol{q}}}$ as the adjacency matrix.  Such analysis would correspond to examining the linear behavior of the perturbation $\boldsymbol{q}'$. 

Assuming the perturbation to be in the form of
$\boldsymbol{q}' = \hat{\boldsymbol{q}}(\boldsymbol{x}) e^{\lambda t}$  
and absence of any forcing $\boldsymbol{f}'$, we arrive at 
\begin{equation}
	\boldsymbol{A} \hat{\boldsymbol{q}}(\boldsymbol{x}) 
	= \lambda \hat{\boldsymbol{q}}(\boldsymbol{x}),
	\label{eq:eig_prob}
\end{equation}
which reduces to an eigenvalue problem.  
This in fact is the global stability analysis \citep{Schmid01,Theofilis:ARFM11} where $\hat{\boldsymbol{q}}(\boldsymbol{x})$ and $\lambda$ represent the stability mode (eigenvector) and the associated growth rate/frequency (eigenvalue), respectively.  
This means that if we consider the adjacency matrix to be the linearized Navier--Stokes operator, the eigenvalue centrality is equivalent to the most dominant instability mode from Eq.~(\ref{eq:eig_prob}).

It is also possible to examine the harmonic response of the flow with sinusoidal forcing input $\boldsymbol{f}' = \hat{\boldsymbol{f}}(\boldsymbol{x}) \exp(i \omega t)$ in which case, we have
\begin{equation}
	\hat{\boldsymbol{q}}(\boldsymbol{x}) 
	= [i \omega \boldsymbol{I} + \boldsymbol{A}]^{-1} 
	\hat{\boldsymbol{f}}(\boldsymbol{x}),
	\label{eq:eig_prob}
\end{equation}
where $\boldsymbol{A}$ is again taken to be $\boldsymbol{L}_{\overline{\boldsymbol{q}}}$.  
In the above equation, $[i \omega \boldsymbol{I} + \boldsymbol{A}]^{-1}$ is called the resolvent operator, which acts as a transfer function between the input $\hat{\boldsymbol{f}}(\boldsymbol{x})$ and output $\hat{\boldsymbol{q}}(\boldsymbol{x})$ for a given frequency $\omega$, and is equivalent to Eq.~(\ref{eq:tf}).  The analysis of the above input-output relationship is known as the resolvent analysis \citep{Trefethen:Science93, Jovanovic:JFM05}.  The most amplified forcing input and the corresponding response can be determined by performing a singular value decomposition of the resolvent operator.  
While network science generally seeks the Katz centrality of the resolvent operator by multiplying the resolvent by $[1,1,\dots,1]^T$ to examine the influence of forcing, it is more insightful to find the forcing and response modes through SVD, yielding the SVD-based broadcast analysis discussed in Section \ref{sec:Katz}.  There is a wealth of insights gained on flow physics from stability analysis and resolvent analysis over the past few decades \citep{Schmid01, Theofilis:ARFM11, Jovanovic:ARFM21, Taira_etal:AIAAJ17, Taira_etal:AIAAJ20}.

\begin{figure}[ht!]
\centering
	\includegraphics[width=0.34\textwidth]{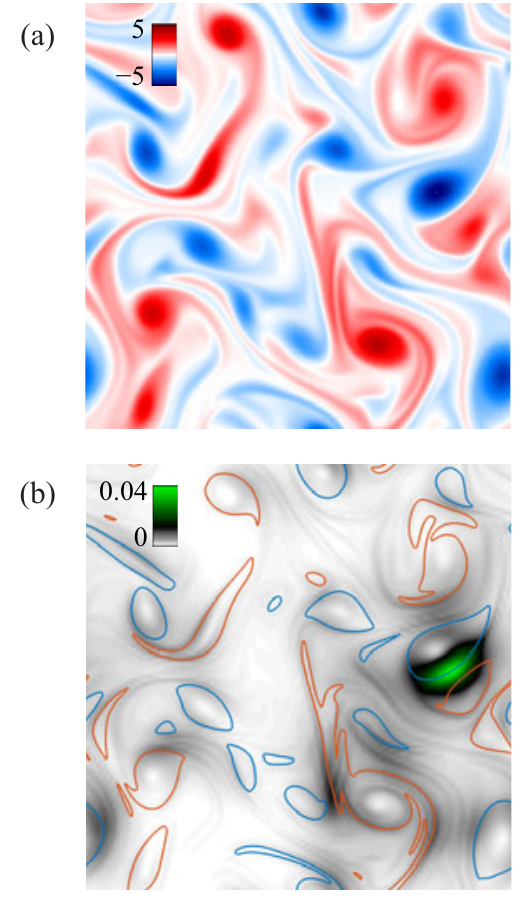}
	\caption{Network broadcast analysis for 2D isotropic turbulence \citep{Yeh:JFM21}.
	(a) Vorticity field $\omega$.  (b) Primary broadcast mode.
	Reprinted with permission from Cambridge University Press.}
	\label{fig:turb_broadcast}
\end{figure}

The linear analysis techniques described above were presented in the continuous time formulation.  It is also possible to perform the analysis in the discrete time formulation.  We can discretize Eq.~(\ref{eq:dyn_f}) in time to arrive at
\begin{equation}
	\boldsymbol{q}'(\boldsymbol{x},t^{n+1}) 
	= \boldsymbol{q}'(\boldsymbol{x},t^n) 
	+ \Delta t \left[ \boldsymbol{L}_{\overline{\boldsymbol{q}}} 
	\boldsymbol{q}'(\boldsymbol{x},t^n) + \boldsymbol{f}'(\boldsymbol{x},t^n) \right].
\end{equation}
This equation was discretized with explicit Euler for demonstration purpose and leads to
\begin{equation}
	\begin{split}
	\boldsymbol{q}'(\boldsymbol{x},t^{n+1}) 
		&= \Delta t [\boldsymbol{I} + \Delta t \boldsymbol{A} 
		    + \dots + \Delta t^{n} \boldsymbol{A}^n] \boldsymbol{f}'(x)\\
		&= \Delta t [\boldsymbol{I} - \Delta t \boldsymbol{A}]^{-1} \boldsymbol{f}'(x)		
	\end{split}
\end{equation}
assuming that we start with zero initial condition and $\boldsymbol{f}'$ being constant in time.  Analyzing the properties of $[\boldsymbol{I} - \Delta t \boldsymbol{A}]^{-1}$ leads to the broadcast analysis \citep{Newman18,Grindrod:PRSLA14}.  This network broadcast analysis has been applied to two-dimensional isotropic turbulence to identify influential structures.  This particular flow is hard to analyze since time-average flow is zero and its flow is chaotically evolving over time.  The work of Yeh et al.~\citep{Yeh:JFM21} further utilized the time-varying broadcast analysis by evaluating the $\boldsymbol{A}(t)$ using automatic differentiation (Fr\'echet derivative) of the nonlinear vorticity transport equation.  Through this analysis, the flow structures that can alter the turbulent flow field efficiently can be identified.  Illustrated in Figure \ref{fig:turb_broadcast} is the primary broadcast mode for example of two-dimensional isotropic turbulence.  Actuation based on the broadcast mode has been shown to effectively modify the mixing dynamics.  As this technique holds similarity to the resolvent analysis but with the ability to examine time-varying dynamics (base flow), it is anticipated to expand the applicability of resolvent based analysis techniques to unsteady flow phenomena.

\subsubsection{Modal interaction network}
\label{sec:modal_interaction}

For unsteady flows, identification of dominant flow structures can aid in the understanding of fundamental dynamics.  
With modal analysis techniques, such energetic and dynamic spatial flow structures can be extracted from the flow field data or the linearized Navier--Stokes operator \citep{Taira:AIAAJ20intro, Taira_etal:AIAAJ17, Taira_etal:AIAAJ20}.  Some well-known modal analysis techniques include the Fourier analysis \citep{pope00, davidson2004turbulence,Canuto88,BoydSpectralMethods}, proper orthogonal decomposition \citep{Berkooz1993,Holmes96,Sirovich:QAM87}, dynamic mode decomposition \citep{Schmid:JFM10,Schmid:ARFM22,Kutz2016book}, stability analysis \citep{Schmid01,Theofilis:ARFM11}, and resolvent analysis \citep{Trefethen:Science93,Jovanovic:JFM05,McKeon2010,Jovanovic:ARFM21}.  
The set of spatial structures (modes) revealed by each of these individual techniques interact among themselves providing richness to the dynamics of unsteady nonlinear flows.  Here, we discuss some of the network-based perspectives used to characterize and model these modal interactions.

Given a set of orthonormal spatial modes $\{ \boldsymbol{\varphi}_1, \boldsymbol{\varphi}_2, \dots, \boldsymbol{\varphi}_r \}$, we can express the state variable $\boldsymbol{q}(\boldsymbol{x},t)$ with a generalized Fourier series of
\begin{equation}
	\boldsymbol{q}(\boldsymbol{x},t) = \sum_{i=1}^r a_i(t) \boldsymbol{\varphi}_i(\boldsymbol{x}),
	\label{eq:gen_fourier}
\end{equation}
where $a_i(t)$ are the temporal coefficients.  For simplicity, we assume the spatial modes to be orthonormal (i.e., $\langle\boldsymbol{\varphi}_i,\boldsymbol{\varphi}_j\rangle = \delta_{ij}$) as in the case of POD and Fourier modes.  
For a governing equation expressed in the form of
\begin{equation}
	\frac{\partial \boldsymbol{q}}{\partial t} 
	= \boldsymbol{L} \boldsymbol{q} 
	+ \boldsymbol{N}(\boldsymbol{q}),
	\label{eq:gov}
\end{equation}
where $\boldsymbol{L}\boldsymbol{q}$ and $\boldsymbol{N}(\boldsymbol{q})$ represent the linear and nonlinear terms, we can insert the generalized Fourier series from Eq.~(\ref{eq:gen_fourier}) to find
\begin{equation}
	\sum_{i=1}^r \frac{d a_i(t)}{d t} \boldsymbol{\varphi}_i(x) 
	= \sum_{i=1}^r a_i(t) \boldsymbol{L} \boldsymbol{\varphi}_i(x) 
	+ \boldsymbol{N}(\sum_{i=1}^r a_i(t) \boldsymbol{\varphi}_i(\boldsymbol{x})).
\end{equation}
Taking an inner product of the terms in the above equation with mode $\boldsymbol{\varphi}_j(x)$, 
\begin{equation}
	\frac{d a_j(t)}{d t} = \sum_{i=1}^r a_i(t) 
	\langle \boldsymbol{\varphi}_j, \boldsymbol{L} \boldsymbol{\varphi}_i\rangle 
	+ \langle \boldsymbol{\varphi}_j, \boldsymbol{N}(\sum_{i=1}^r a_i(t) \boldsymbol{\varphi}_i ) \rangle,
	\label{eq:galerkin}
\end{equation}
where $j = 1, \dots, r$. 
In the above derivation, we have used the orthonormal property of the modes.
If the modes are non-orthogonal, the left-hand side term will carry a mass matrix.  

\begin{figure*}[ht!]
\centering
	\includegraphics[width=0.8\textwidth]{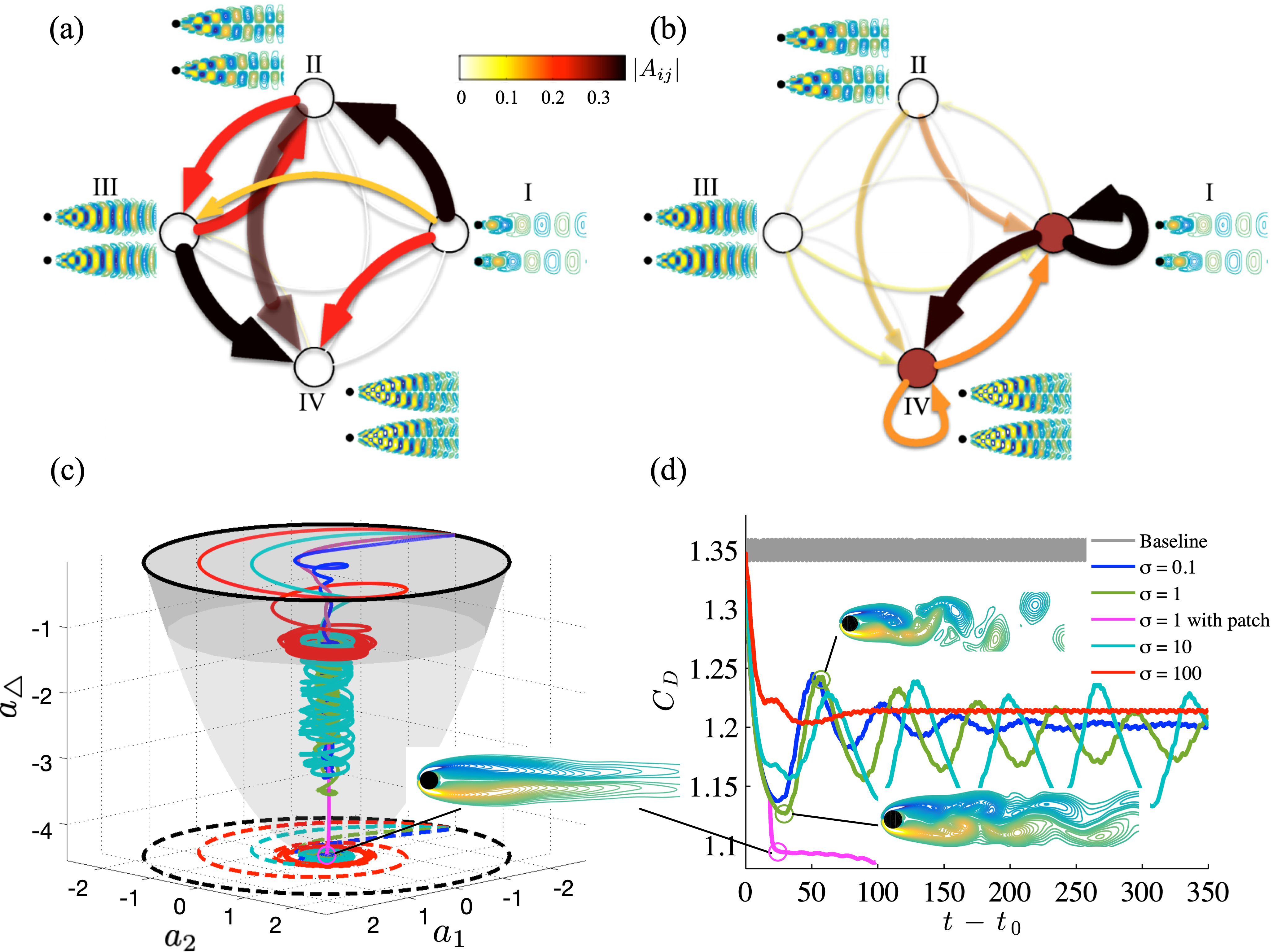}
	\caption{Modal interaction network for laminar flow over a cylinder \citep{nair2018network}: (a) Perturbation-induced interactions between the first 8 POD modes of the flow, (b) altered interactions with optimal control added to the modes indicated by the red filled circles. Also shown are (c) the networked dynamics along the parabolic manifold \citep{brunton2015closed} with different LQR gains and (d) the corresponding drag coefficients with and without control.
	Reprinted with permission from the American Physical Society.
	}
	\label{fig:network_oscillators}
\end{figure*}

Note that the original governing equation (\ref{eq:gov}) describes the evolution of a high-dimensional variable $\boldsymbol{q} \in \mathbb{R}^n$ or $\mathbb{C}^n$ ($n \approx$ the number of grid points times the number of state variables).  On the other hand, Eq.~(\ref{eq:galerkin}) describes the dynamics with only $r$ equations for the temporal coefficients $\{ a_1, \dots, a_r\}$, which is significantly reduced (i.e., $r \ll n$).  For this reason, Eq.~(\ref{eq:galerkin}) is referred to as a reduced-order model.  The specific approach used to derive this model is known as the Galerkin projection \citep{Holmes96,Noack:JFM03,Noack11,Ahmed:PF21}.  This formulation is also synonymous with the wavespace description of the governing equation for Fourier methods \citep{davidson2004turbulence,Canuto88,BoydSpectralMethods}.

From a network perspective, we observe that Eq.~(\ref{eq:galerkin}) describes the interactions amongst modes $\{ \boldsymbol{\varphi}_1, \dots, \boldsymbol{\varphi}_r \}$.  Hence, we can regard this equation as a basic dynamical representation for the modal interaction network, in which the nodes are the modes and the edges are the interactions between them.  Here, matrix $\left< \boldsymbol{\varphi}_j, \boldsymbol{L} \boldsymbol{\varphi}_i \right>$ serves as the adjacency matrix.  
One important question is how we interpret the nonlinear term $\left< \boldsymbol{\varphi}_j, \boldsymbol{N}\left(\sum_{i=1}^r a_i(t) \boldsymbol{\varphi}_i\right) \right>$ in Eq.~(\ref{eq:galerkin}).  If we are concerned with the linearized dynamics about some equilibrium state, this nonlinear term can be small in magnitude and be neglected.  When that is not the case, the nonlinear term must be considered carefully as we discuss below with some examples.

The modal interaction network has been used to model and control the unsteady wake behind a circular cylinder \citep{nair2018network}.  In this work, POD mode pairs having equal kinetic energy content constitute a complex mode, serving as the network node.  This formulation yields an oscillator network as the complex mode represents a limit cycle in the absence of perturbations, as shown in Figure \ref{fig:network_oscillators}.  This modal interaction network captures kinetic energy transfer modeled over a diffusive network for which the graph Laplacian represents the linear interactions.  In this case, the nonlinear term was also approximated by the linear model, similar to how dynamic mode decomposition and the Koopman formalism incorporate nonlinearly into the linear representation \citep{Koopman1931pnas,Schmid:JFM10,Kutz2016book}.
The graph Laplacian was determined through regression-based on time series from perturbed flow simulations. 
As the node is complex-valued, the adjacency matrix for this case possesses real and imaginary components, which capture the energy transfer and the phase advancement/delay effects, respectively.
This modeling effort also revealed that the first POD pair serves as the leader node and the remaining nodes act as followers \citep{Mesbahi10}.
Furthermore, feedback control based on this networked POD oscillator model was implemented to suppress wake oscillations for significant drag reduction \citep{nair2018network}, as presented in Figure \ref{fig:network_oscillators}.

The use of the Galerkin projection model was also considered by Rubini et al.~\citep{Rubini:JFM20}, in which POD modes were used as the basis to describe unsteady open cavity flow through a sparse model based on $L_1$ optimization.  In this case, linear and nonlinear interactions were retained with the average interactions were analyzed in detail.  As demonstrated in this work, the sparse model provides a pathway to extracting dominant interactions that are important for the overall networked dynamics.

\begin{figure}[ht!]
\centering
	\includegraphics[width=0.43\textwidth]{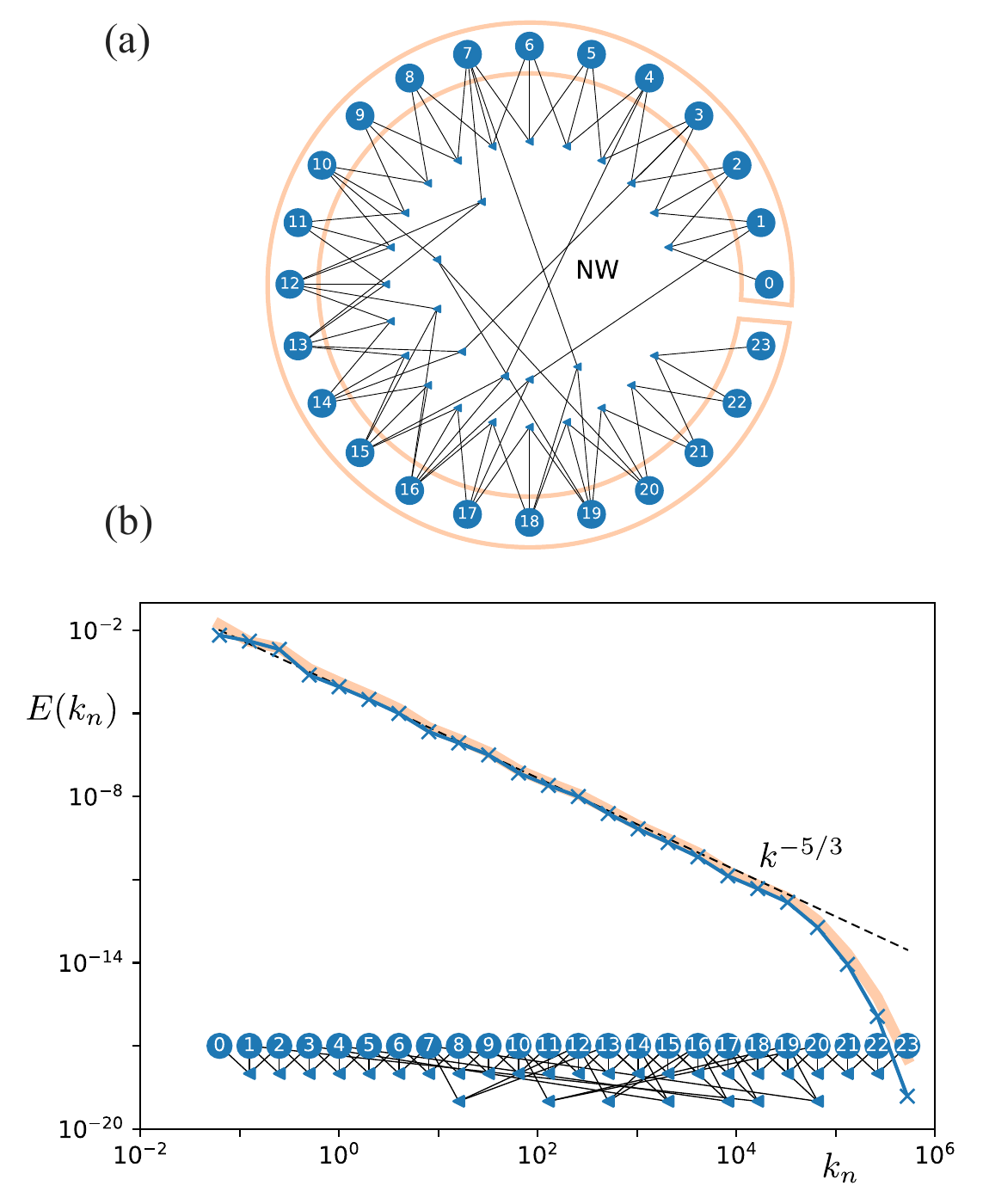}
	\caption{Turbulence interactions modeled by a Fourier mode network \citep{Gurcan:PhysD21}. (a) Newman-Watts small-world model for shell modeling with non-local interactions.
	(b) Kinetic energy spectrum predicted by the small-world shell model. 	
	Reprinted with permission from Elsevier.}
	\label{fig:gurcan_network}
\end{figure}

Fourier modes in wavespace can also be used to describe the interactive dynamics in fluid flows, which has been utilized to examine turbulence and establish Fourier spectral methods \citep{davidson2004turbulence,Canuto88}.
Recently, G\"urcan used the Fourier representation of turbulent flow to establish a modal interaction network \citep{Gurcan:PhysD21}, as illustrated in Figure \ref{fig:gurcan_network}.  In this work, the velocity variable is discretized with Fourier modes that are collected over logarithmically spaced wavenumbers, constituting shells \citep{Biferale:ARFM03}.  The triadic nonlinear interactions among these shells are captured through the introduction of triadic interaction nodes.  
These triadic interaction nodes are connected with shell nodes over a bipartite network. 
This network was initially provided with regular connections and then modified through rewiring based on Watts--Strogatz \citep{Watts:Nature98} and Newman--Watts algorithms \citep{Newman:PRE99}.  
For an example of three-dimensional isotropic turbulence, it was shown that dynamically rewiring the network enables this model to exhibit the expected kinetic energy spectra.  
Moreover, intermittent dynamics appeared more frequently for networks with a larger ratio of nonlocal to local connections.  
This refreshing perspective to characterize modal interactions has the potential to describe important nonlinear interactions in a variety of laminar and turbulent flows.

\subsubsection{Lagrangian transport network}
\label{particleproximity}

We can track the motion of tracer particles advected by the flow to build proximity or similarity networks. Seeding the flow with suitable tracer particles is typical in particle-image velocimetry (PIV) experiments \citep{melling1997tracer}. By viewing the particles or groups of particles to be nodes and edges between them based on the relative proximity or similarity of their trajectories, a flow field network can be characterized.

One way to quantify the edges between particles is to consider the spatial proximity of their trajectories with each other. The proximity can either be assessed over a finite time window $T$ or at each time instant, generally leading to a time-varying network. The link between the particles can be defined as
\begin{equation}
    w_{ij} = 1/\overline{r_{ij}},~~~\text{with~} r_{ij}(t) = \|\boldsymbol{r}_i(t) - \boldsymbol{r}_j(t)\|_2 
\end{equation}
where $\overline{r_{ij}}$ denotes the time-average distance between the two particle trajectories \citep{hadjighasem2016spectral}. The link between particles $i$ and $j$ is stronger if their trajectories are closer together. This concept of proximity network is similar to that of the vortex interaction network but without the influence of induced motion on each other.  

\begin{figure}[ht!]
\centering
	\includegraphics[width=0.48\textwidth]{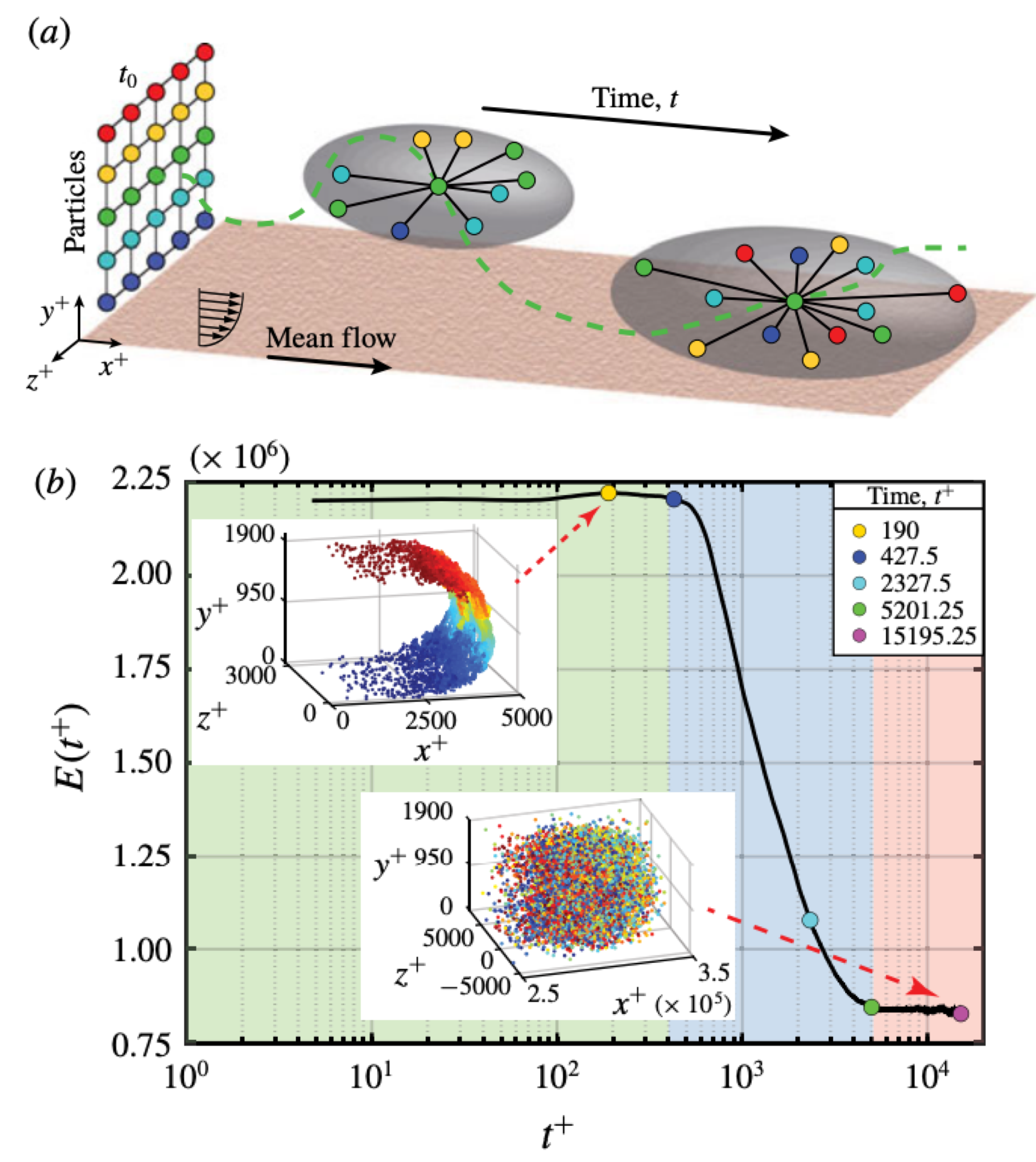}
	\caption{Turbulent channel flow Lagrangian network analysis  \citep{iacobello2019lagrangian}: (a) Schematic of analysis setup where particles are initially released from a uniformly spaced grid at $x^{+} = 0$. The colored spheres represent particles and black lines indicate the connections. (b) Total number of connections $E(t^+)$, as a function of time in turbulent channel flow. Reprinted with permission from Cambridge University Press.}
	\label{fig:particlenetwork0}
\end{figure}

There has also been an alternative approach considered by forming an undirected, unweighted network.  In this formulation, $A_{ij} = 1$ if and only if both particles $i$ and $j$ are found within a ball of radius $\epsilon$ at one or more time instances \citep{padberg2017network}.  The network in this case is sensitive to the choice of $\epsilon$. A generalization of the above approaches was pursued by \citet{iacobello2019lagrangian} for turbulent channel flow as shown in Figure \ref{fig:particlenetwork0}(a). The total number of network connections $E = \frac{1}{2} \sum_{i,j} w_{ij}$ established at each non-dimensional time instant $t^{+}$ of the evolution of turbulent channel flow is shown in Figure \ref{fig:particlenetwork0}(b). This metric identifies three flow regimes, namely, an initial regime where particles are only connected if they are near one another, a second regime where the wall-normal mixing tends to separate the initially nearby particles leading to a sharp drop in total network connections, and a final regime where a balanced Taylor asymptotic state of transversal mixing has been established.

\begin{figure}[ht!]
\centering
	\includegraphics[width=0.42\textwidth]{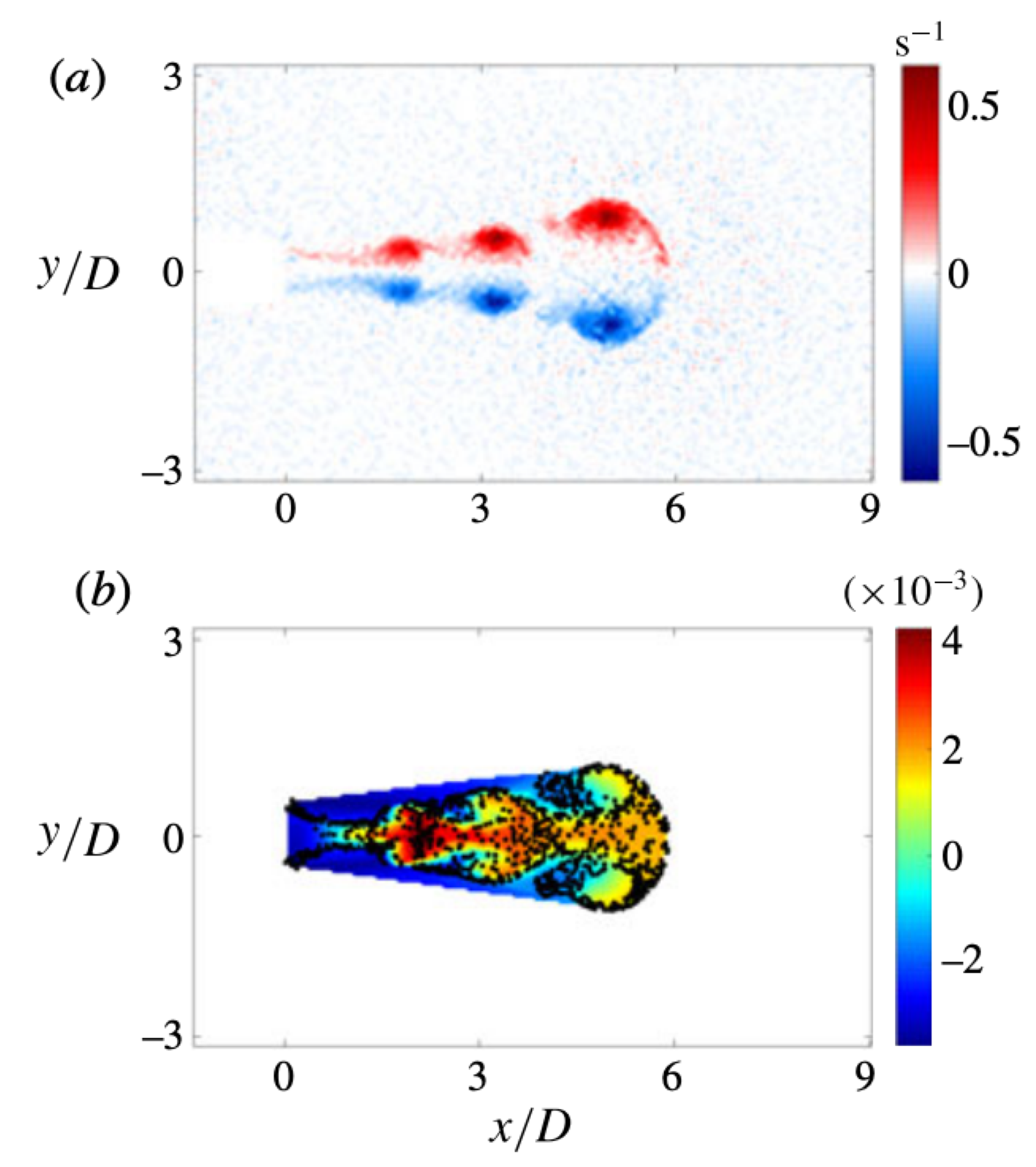}
	\caption{Vortex ring coherent structure coloring \citep{schlueter2017jfm}: (a) Vorticity field at $t^{*} = 10.2$ and (b) coherent structure coloring over the time interval $t^{*} = [8.0, 10.2]$ with 1200 particles introduced at the nozzle plane. Reprinted with permission from Cambridge University Press.}
	\label{fig:particlenetwork}
\end{figure}

Another way to define the edge weights is to consider the similarity in the particle kinematics. For such a similarity-based network, if the pathlines of particles are similar, then the link between them is stronger regardless of their spatial proximity. The edge weight can be defined by the standard deviation of the distance between two-particle trajectories over a time horizon $T$ normalized by the average distance between the particle trajectories over the same time horizon \citep{schlueter2017jfm} as
\begin{equation}
	A_{ij} = \frac{1}{\overline{r_{ij}}} \left[ \frac{1}{T}\sum_{k = 0}^{m-1} \left(\overline{r_{ij}} - r_{ij}(t_k)\right)^2\right]^{1/2}.
\end{equation}
Using the above adjacency matrix, a normalized Laplacian matrix $\mathcal{L} = \boldsymbol{D}^{-1}\boldsymbol{L}$ for the similarity-based network can be evaluated. The eigenvector corresponding to the maximum eigenvalue of $\mathcal{L}$ identifies regions of the flow that are coherent \citep{schlueter2017jfm, husic2019simultaneous}. Such a field is shown for the PIV dataset of a vortex ring in Figure \ref{fig:particlenetwork}, where the flow is seeded with $1200$ particles introduced at the nozzle exit plane \citep{schlueter2017jfm}. 

The transport of particles can also be examined by discretizing the fluid domain into $N$ discrete fluid boxes $B_j$, where $j = 1, 2, \cdots, N$. Each box can be considered as a node and a directional link between the boxes can be established as the function of the number of particles transported from one box to another \citep{ser2015flow}. For a particular initial time $t_0$ and elapsed time $\tau$, the weighted and directed adjacency matrix can be defined as 
\begin{equation}
    A_{ij} = \frac{\textsf{No.~of particles advected from $B_j$ to $B_i$}}{N_j},
\end{equation}
where $N_j$ is the initial number of particles in $B_j$. The matrix summarizes the final location of the particles at time $t_0 + \tau$. \citet{ser2015flow} analyzed the Mediterranean sea surface data using this approach and found that the out-degree of the nodes is related to the fluid stretching while the in-degree quantified the fluid dispersion. Also, network-entropy functions provided good estimates of the coarse-grained finite-time Lyapunov exponents of the flow \citep{boffetta2002predictability}.  We also note that this approach has some similarities to the cluster-based network, which is discussed later in Section \ref{sec:cluster_network}.

\subsection{Sensor-based networks}
\label{subsec:tsn} 

In this section, we present sensor-based networks inspired by time-series analysis and dynamical systems theory.  These approaches are amenable to network characterizations of flows for where only limited sensor measurements are available. This is typical in experiments where the collection of spatially resolved flow fields may be difficult. Taking advantage of phase-space techniques and time-delay embedding, the interactive physics of the flows can be extracted by the following sensor-based network approaches.

\subsubsection{Visibility graph}
\label{sec:visibility}

Visibility graph converts univariate time-series measurements into a network. The original version of this approach \citep{lacasa2008time, lacasa2012time} takes the individual temporal measurements as nodes and the presence or absence of edges is established if the observations are directly visible to each other, i.e., if a straight line connects the observations. More formally, nodes $i$ and $j$ are connected if any sensor measurement $y_k = y(t_k)$ that lie between the measurements $y_i = y(t_i)$ and $y_j = y(t_j)$ satisfy the relation
\begin{equation}
    \frac{y_k - y_j}{t_j- t_k} <  \frac{y_i - y_j}{t_j - t_i}.
\end{equation}
A less cumbersome construction is the horizontal visibility criterion \citep{luque2009horizontal} which states that nodes $i$ and $j$ are connected if $y_k < \text{min}(y_i, y_j)$ for all $t_k$ within $t_i \le t_k \le t_j$. The main advantage of the visibility graph analysis is that it is able to distinguish between periodic signals and random signals, with the former yielding a regular network and the latter an exponential random network. The visibility graph approach has been successfully used to discriminate chaotic time-series from uncorrelated random measurements \citep{luque2009horizontal}.  It can also be extended to bivariate time-series \citep{zou2014long}. For a comprehensive review on visibility graphs, we refer readers to \citet{nunez2012visibility} and \citet{zou2019complex}.

\begin{figure}[ht!]
\centering
	\includegraphics[width=0.45\textwidth]{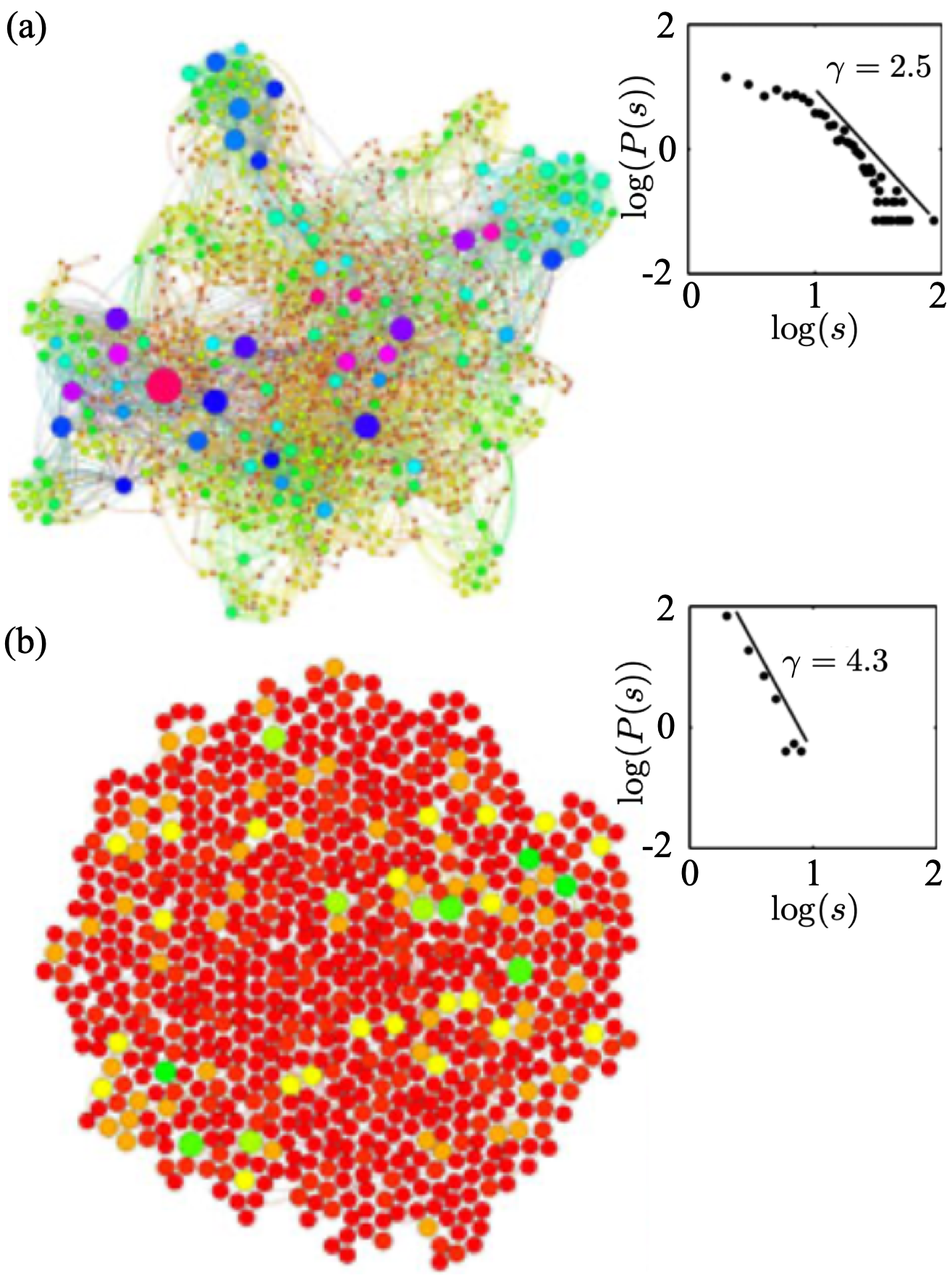}
	\caption{Degree distributions and network structure of a thermoacoustic system using visibility graph analysis \citep{murugesan2015combustion} for (a) $Re = 2.2 \times 10^4$ and (b) $2.8 \times 10^4$. Reprinted with permission from Cambridge University Press.}
	\label{fig:combustionnetwork}
\end{figure}

\citet{murugesan2015combustion} converted acoustic pressure time-series measurements from a thermoacoustic system into a visibility graph as shown in Figure \ref{fig:combustionnetwork}. Much like the scale-free behavior for the turbulence in \S\ref{vorticalnetwork}, scale-free behavior is observed for the combustion noise visibility graph with a power-law exponent in the range $\gamma =  2.5 - 2.7$, corresponding to combustion in turbulent flows as seen in Figure \ref{fig:combustionnetwork}(a). Upon increasing the Reynolds number of this system, the emergence of combustion instability is observed with the presence of intermittent bursts of periodic oscillations as shown in Figure \ref{fig:combustionnetwork}(b). As expected, the regularity of the degree of the nodes in the visibility graph increases with periodic behavior.  Through the imposition of constraints on the visibility graphs, the online predictability of the combustion dynamics has been shown to improve \citep{gotoda2017characterization}.

The visibility graph was also used to examine the spatio-temporal dynamics of a turbulent co-axial jet by \citet{kobayashi2019spatiotemporal}. This flow transitions from a stochastic to a chaotic state as the jet evolves downstream. The analysis highlighted the small-world and scale-free properties of the network in the stochastic and chaotic regimes, respectively. The application of visibility graphs was extended to turbulent premixed flames on a model gas turbine combustor where the flame front was considered as the nodes of the network \citep{singh2017network}. The hub nodes of the network were found to exist in the high-curvature regions of the front. The change in the properties of fully-developed turbulent channel flow in the wall-normal direction \citep{iacobello2018visibility, iacobello2021large} and convective surface layer turbulence \citep{chowdhuri_iacobello_banerjee_2021} was also investigated using visibility graph analysis. The visibility graph analysis is also capable of extracting the temporal structure of extreme events and their intensity in experimental measurements of atmospheric turbulent boundary layer flows \citep{iacobello2019experimental}. 

\subsubsection{Recurrence network}
\label{sec:recurrence}

Instead of considering each sensor measurement as the nodes of the network, the sensor measurements can be aggregated as $\boldsymbol{r}_i = (y_i, y_{i+\tau},\cdots,y_{i + \tau(m-1)})$. Here, $\tau$ is the time delay between each measurement, and $m$ is the number of measurements aggregated. In a similar manner, other measurements replacing $i$ with $j$ can be aggregated to form $\boldsymbol{r}_j$. These aggregated variables $\boldsymbol{r}_i$ and $\boldsymbol{r}_j$ are called phase-space vectors and together they form the phase-space of the system. 
Taken's embedding theorem \citep{takens1981detecting} gives us the conditions in which a dynamical system behavior can be reconstructed from the phase-space vectors. In fact, each possible state of the unsteady fluid flow can be represented as a unique point in the corresponding phase space.  Different phase-space vectors can be constructed based on the choice of the embedding dimension $m$ \citep{kennel1992determining, cao1997practical} and time-delay $\tau$ between measurements \citep{fraser1986m}. 

If the state of the fluid flows is recurrent, the corresponding phase-space trajectories are closer than a particular threshold. This recurrence behavior in the phase space is encoded with the recurrence matrix given by 
\begin{equation}
    R_{ij} = \Theta (\epsilon - ||\boldsymbol{r}_i  - \boldsymbol{r}_j||_2),
\end{equation}
where $\Theta$ is the Heaviside step function and $\epsilon$ is a chosen threshold value. This matrix takes $R_{ij} = 1$ if the distance between the corresponding trajectories is less that $\epsilon$ and $R_{ij} = 0$ otherwise. The diagonal entries of the recurrence matrix are always unity. Subtracting the effect of the self-loops in the recurrence matrix, an adjacency matrix can be defined from the recurrence matrix as $A_{ij} = R_{ij} - \delta_{ij}$, where $\delta_{ij}$ is the Kronecker delta.

Instead of selecting a threshold $\epsilon$, the $k$-nearest neighbors to a node can be chosen to define the network. Here, a directed edge is introduced from node $i$ to the $k$-nearest neighbors with the shortest mutual distances with phase space vector $\boldsymbol{r}_i$ \citep{shimada2008analysis, donner2011recurrence}. If the in-degree $s_i^\text{in} \gg k$, then node $i$ is located in a densely populated region of the attractor of the flow. The nearest neighbors can also be adaptively defined yielding undirected recurrence networks \citep{xu2008superfamily}.

Previous efforts have constructed recurrence networks from experimental acoustic pressure time-series measurements at different equivalence ratios for a combustor. The characteristic path length and betweenness centrality measures of the networks provide early warning indicators for transitions from combustion noise to thermoacoustic stability \citep{godavarthi2017recurrence}. Moreover, the presence of a small-world network structure for a recurrence network has been reported for the combustion states close to blowout \citep{gotoda2017characterization}. 

Alternative to establishing the connections based on recurrence, \citet{zhang2006complex} mapped the connections between each cycle of pseudo-periodic time series by the average correlation coefficient between them. The application of cycle networks to a time-series of acoustic measurements from a thermo-acoustic system helped distinguish the different dynamical states and characterize the stability of dynamically invariant periodic orbits in phase space \citep{tandon2021condensation}.

\subsubsection{Cluster transition network}
\label{sec:cluster_network}

The individual trajectories of the particles were tracked to construct Lagrangian transport networks in section \ref{particleproximity}. 
A complementary approach is to track the time evolution of an ensemble of such trajectories in the appropriate coordinates (phase space) of the system \citep{li1976finite, cvitanovic2005chaos}. In fact, the time evolution of the probability densities of these phase-space trajectories is governed by the linear Perron--Frobenius operator associated with the Liouville equation \citep{lasota2013chaos}. 

A finite-rank approximation to the Perron--Frobenius operator can be obtained through Ulam's conjecture \citep{bollt2013applied, Kaiser:JFM14}. Consider the measurements $y_{k} = y(t_k)$ in the phase space such that 
\begin{equation}
y_{k+1} = {F}(y_{k}),
\end{equation}
where ${F}$ is some discrete mapping function.  Discretizing the phase space into $N$ boxes $\{B\}_{i = 1}^N$, the matrix approximation to the Perron--Frobenius operator is given as
\begin{equation}
P_{ij} = \frac{\text{card}(y_{k}| y_{k} \in B_j~\text{and}~y_{k+1} \in B_i)}{\text{card}(y_{k} \in B_i)},
\label{Ulam}
\end{equation}
where $\text{card}(\cdot)$ represents the cardinality of the set or the number of elements in that set.  Matrix $P_{ij}$ represents the fraction of observations in box $B_j$ that transfers to $B_i$ after one iteration of the discrete map. The dominant eigenvector of this matrix approximates the invariant probability distribution of the Perron--Frobenius operator \citep{bollt2013applied}.  

\begin{figure*}[ht!]
\centering
	\includegraphics[width=0.7\textwidth]{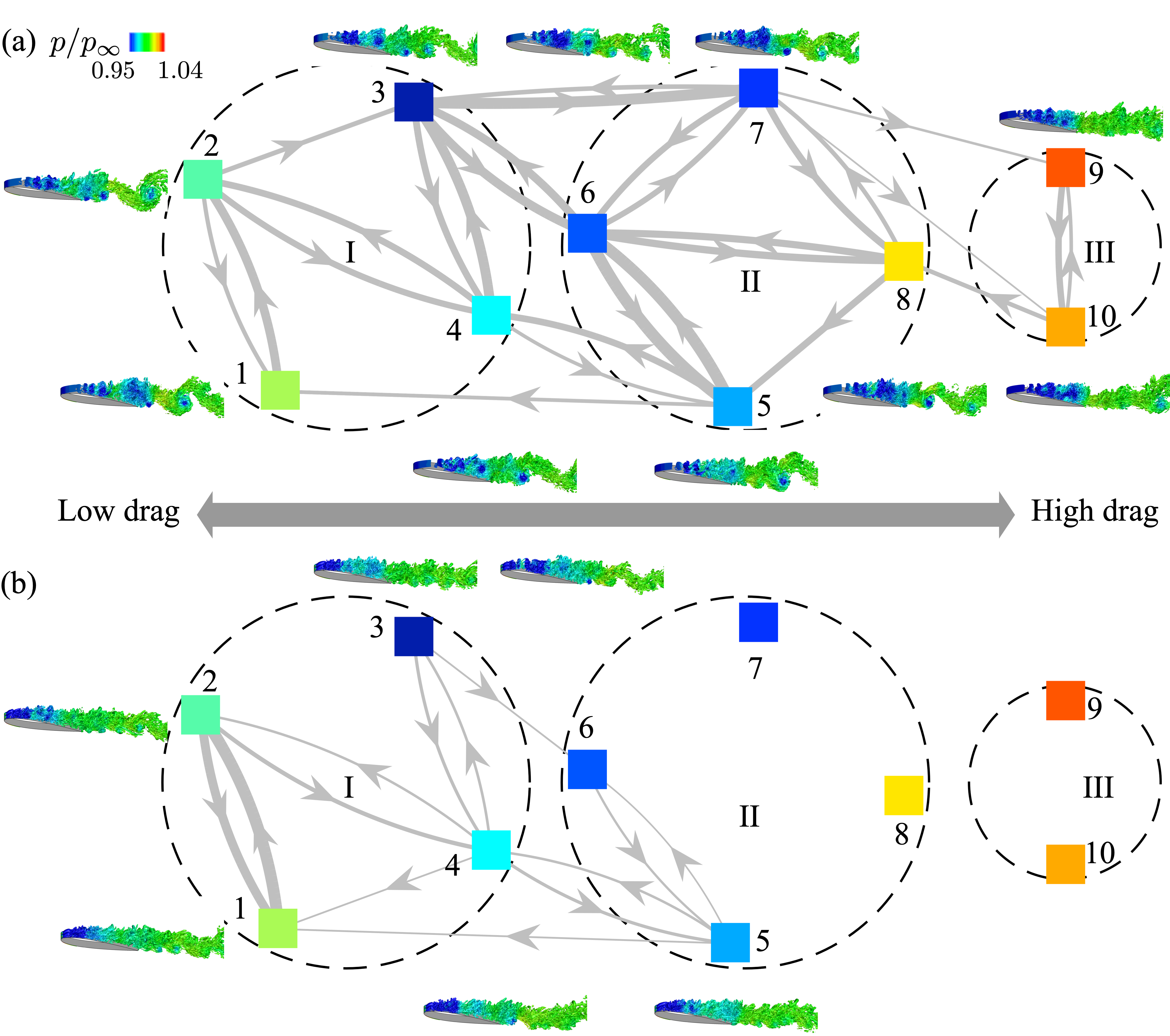}
	\caption{(a) Baseline and (b) controlled transition network for turbulent separated flow over an NACA 0012 airfoil at post-stall angle of attack of $9^\circ$ \citep{Nair:JFM2019}.
	Reprinted with permission from Cambridge University Press.
	}
	\label{fig:clusternetwork}
\end{figure*}

Recently, \citet{Kaiser:JFM14} generalized Ulam's method. Instead of discretizing the phase space based on the geometrical proximity of measurements, a more flexible strategy of cluster analysis was adopted.  The cluster analysis relies on the choice of a distance metric and clustering algorithm. The standard Euclidean distance or any other measure of similarity between measurements could be used, akin to particle similarity networks. The measurements are then partitioned into clusters using a clustering algorithm. One such algorithm is {\it k-means}, which partitions the measurements into clusters $\{\mathcal{C}_k\}_{k=1}^K$ such that the inner-cluster similarity is maximized, while inter-cluster similarity is minimized \citep{rokach2005clustering}. The likelihood that the measurements remain in cluster $\mathcal{C}_k$ is defined as $q_k = {n_k}/{M}$, where $n_k$ is the number of measurements in each cluster and $M$ is the total number of measurements. 
The probability of transition of observations from cluster $\mathcal{C}_j$ to $\mathcal{C}_i$ in one forward time-step is given by 
\begin{equation}
P_{ij} = {n_{ij}}/{n_j},
\label{transition_matr}
\end{equation}
which is analogous to the transition matrix in Eq.~({\ref{Ulam}}) using Ulam's method. Both these transition matrices are non-negative weighted directed networks with $\sum_{j=1}^K P_{ij} = 1$ to preserve the normalization of probabilities. The diagonal elements of the network indicate the number of within-cluster transitions. 
\citet{fernex2021cluster} modified the transition matrix in Eq. (\ref{transition_matr}) such that the inner-cluster residence probability is removed, producing a dynamic propagator model for the cluster states.

Since its inception, cluster transition network analysis has been applied to  two-dimensional mixing layer \citep{Kaiser:JFM14, li2021cluster}, two-phase flows \citep{ali2020cluster}, wind turbine wake \citep{ali2021cluster} and wake of a high-speed train \citep{osth2015cluster}. \citet{schmid2018description} used cluster analysis to detect burst events in a model of turbulent wall-bounded shear flows. The cluster transition network for turbulent separated flow over a NACA 0012 airfoil at a post-stall angle of attack are shown in Figure \ref{fig:clusternetwork} (a). Here, k-means clustering is performed on measurements of the lift coefficient $C_L$, its time-derivative $\dot{C}_L$, and the drag coefficient $C_D$ \citep{Nair:JFM2019} to partition the phase-space data into 10 clusters. The representative flow field for the centroid of each cluster is shown. Modularity maximization of the transition matrix $P_{ij}$ reveals three communities corresponding low-drag, intermediate drag and high-drag states of the turbulent flow. As seen in Figure \ref{fig:clusternetwork} (a), the transitions from cluster 3 to 6 and 7 lead the dynamics to high-drag states. By introducing adaptive control inputs at each cluster centroid, the transitions can be modified such that the flow primarily oscillates between the clusters in the low-drag community, as seen in Figure \ref{fig:clusternetwork}(b).

The choice of the variables (features) used for cluster analysis plays an important role in capturing the critical transitions and characterizing the interactions. In the work by \citet{Kaiser:JFM14}, observations over the entire flow field were used to construct a cluster-based reduced-order model. If the objective is to control the flow using limited measurements, then it is preferable to use the features that also define the objectives for flow control \citep{Nair:JFM2019}.


\section{Outlook}
\label{sec:outlook}

Here, we discuss a few topics that have implications for the applicability of network science to fluid mechanics.  In particular, we present below our perspectives on weighted networks, multi-layer networks, hypergraphs, governing equation discretizations, graph-theoretic flow control, computational algorithms, and machine learning, in which advancements may significantly expand the horizon of network-based analysis of fluid flows.

\subsection{Weighted networks}

Network science originally was been established primarily with unweighted graphs, leading to the developments of algorithms aimed at unweighted connections \citep{Bollobas98,Newman18,Chung97}.  However, in continuum mechanics, the values of weights are on a continuous scale, requiring the use of a weighted network \citep{Barrat:PNAS04}.  Moreover, the concept of signed edges is necessary to describe the orientation of vectors.  The extensions of available network science toolsets for weighted networks are sometimes still under development and await progress.  Studies of physical interactions can benefit from progress in network science or research inputs from fluid mechanics in this regard.

\subsection{Multi-layer networks}

There are scenarios where multiple networks are called for to analyze multi-physics problems. Such a system can be concisely represented by a multi-layer network formalism \citep{Bianconi18,Kivela:JCN14,Aleta:ARCMP19}. This opens avenues for reducing the complexity of each system while retaining the important inter-system connections. This should have implications for multi-body wakes, fluid-structure interactions, aeroacoustics, biological propulsion, wing-disturbance interaction, and chemically-reacting flows. 

\subsection{Hypergraphs}

In this paper, the focus of the discussion was placed on interactions between two elements.  However, the concept of interactions can be generalized beyond two elements.  Hypergraphs provide such a formulation to describe multi-node interactions (beyond two elements) on what is known as the hyperedge \citep{Newman18,Dorogovtsev10}.  As triadic interactions serve a critical role in laminar \citep{Noack11,Callaham:JFMXX} and turbulent flows \citep{Hinze75,pope00,davidson2004turbulence}, hypergraphs may facilitate its theoretical analysis.  As a matter of fact, this type of graph has a close connection to the tensorial representation of the adjacency matrix.  Hypergraphs can also be represented by the composition of two-dimensional graphs through the use of bipartite networks.  Moreover, the bispectral analysis appears promising in revealing important nonlinear interactions in wavespace \citep{Kim:IEEE79, schmidt2020bispectral}.

\subsection{Discretization}

At the moment, there is no formal procedure to translate governing equations for fluid flows into networked dynamical systems suitable for graph-theoretic analysis. The discretization framework needs to ensure that the derived networked dynamical systems honor conservation properties of the original flow dynamics \citep{gustafson1985graph, amit1981application, hall1992network, Pozrikidis14}.  
Extraction of networked dynamics in an interpretable manner while preserving physical properties \citep{loiseau2018constrained} may hold the key to not only modeling the interaction dynamics but also controlling the dynamics of unsteady fluid flows.

\subsection{Computational algorithms}

In fluid mechanics, the degrees of freedom of the system can be very large, especially for turbulent flows.  Generally, the degrees of freedom (dimension of the discrete state variable) is the product of the number of grid points and the number of unknown variables.  In some cases, that degree of freedom can be significantly larger than what is generally encountered in standard network science.  It should be noted that the adjacency matrix is the size of such degree squared.  Moreover, networks for fluid flows can be much denser due to the broadband nature of interactions \citep{Taira2016jfm,MGM:JFM21}.  This means that the computational algorithms and memory requirements can be taxing if used naively \citep{bai2019randomized}.  Capitalizing on expertise from computational fluid dynamics \citep{Lomax01,Ferziger02,Kajishima17} and data-driven techniques \citep{Brunton2016pnas,Halko2011siamreview,Brunton2019book,Watt20} to advance algorithms for network science will likely allow for the interaction-based studies of large-scale/high-Reynolds-number turbulent flows.

\subsection{Graph-theoretic flow control}

By establishing a network-based description of the dominant flow dynamics, it is possible to design graph-theoretic feedback control \citep{Rahmani:SIAMJCO09, Mesbahi10, nair2018network}.  With a perspective that combines rigorous control theory and graph theory, we can consider the connectivity of interacting fluid elements to develop control strategies with a variety of cost functions.  There are frameworks to perform consensus agreement and multi-agent control, which should be relevant for a range of fluid flows.

\subsection{Machine learning}

There is a close relationship between the ideas presented in this paper and machine learning. 
With the surge in utilizing machine learning, we have seen novel machine learning approaches to analyze and model fluid flows.  
There are two broad categories of machine learning: unsupervised and supervised machine learning techniques \citep{brunton2019machine}. Unsupervised techniques include methods such as clustering which serves as the basis for community detection \citep{Porter:AMS09}. On the other hand, supervised techniques include regression which serves an important role in the data-driven inference of network structures and dynamics \citep{yamanishi2004protein, vert2005supervised, casadiego2017model, newman2018estimating, mauroy2019koopman}. 

Furthermore, there is some overlap between the concept of network structures and those used by neural networks \citep{scarselli2008graph, xu2018powerful, sanchez2020learning, maulik2020probabilistic}. In the construct of neural networks, graph theory is leveraged to establish and modify network architectures, e.g., bipartite graphs, graph pruning \citep{musslick2016controlled, prabhu2018deep, zhou2020graph}.
However, the structures of neural networks to be used by machine learning remains to incorporate physical insights based on the concepts presented herein. There are ongoing efforts to develop physics-inspired networks by embedding physical constraints, laws, and insights into the network structure or loss function \citep{raissi2019physics, mao2020physics, fukami2019super, fukami2021super}. Moreover, achieving interpretability of neural networks is another area that can benefit from future investigations \citep{molnar2020interpretable}. We foresee that understanding the relationship between fluid-flow networks discussed in this paper and neural networks used in machine learning would expand the envelope of both areas of research. 


\section{Conclusions}
\label{sec:conclusions}

This paper offered a brief introduction to network science and reported on the current status of network-inspired analysis, modeling, and control for a variety of unsteady fluid flows.  
With the general formulation of network science allowing for nodes and edges to be assigned to different elements and physics of fluid flows, different aspects of flow interactions can be highlighted to offer insights into networked flow dynamics and structures.   
In this survey, we presented successful applications of network science to study bluff-body flows, turbulence, chemically reacting flows, and flow control.
Here, we have broadly categorized the formulations into the flow field networks and sensor networks.

The former networks used spatial flow information to define the nodes and edges.  Vortical networks in Lagrangian and Eulerian perspectives were used to capture interactions among vortices.  We also showed that there are close ties between linear dynamical systems and network-based analysis.  Modal interaction networks that are founded on modal basis functions were also discussed.  The resulting models have enabled the application of feedback control to modify the networked fluid flow dynamics.  Proximity networks were also introduced where particle trajectories were studied to capture flow features and scale similarity characteristics to classify flow dynamics.

The latter approach converts time-series measurements from flow sensors to networks. Visibility analysis characterized the network connections based on the mutual visibility of the measurements. Recurrence and cluster transition networks extracted the recurrence behavior and transitions in the embedded trajectories in the phase space of the flow, respectively. The resulting networks enable early warning indicators for critical flow transitions and allowed for modeling and adaptive control of complex turbulent flow. 

Exciting progress is being made in network science and data science that has the potential to enable the analysis of complex dynamics of various flows.  Algorithms and theories related to weighted networks, multi-layer networks, and hypergraphs are developed and utilized for complex networks, which should have implications for fluid flow analysis, especially for those with multi-physics and strong nonlinearities.  Moreover, graph-theoretic control techniques that take advantage of the knowledge of connectivities to control multiple attributes of the system can be very useful for active flow control endeavors.  Network science is also the beneficiary of the concepts and methods that are bursting out of the data science and machine learning communities.  The computational innovations from these areas are enabling the analysis of large-scale networks with complex structures in an efficient manner.  The incorporation of these novel concepts and toolsets will further advance the capability of network-inspired study of fluid flows. 

With rapid developments in the field of network science, analysis, modeling, and control of networked dynamics have become possible for highly complex large-scale problems.  Network-based analysis has deepened the understanding of fundamental phenomena in a wide range of disciplines, including biology, social science, computer science, and engineering.  It continues to advance the understanding of brain functions and diseases as well as how society reacts to information or diseases.  Given the unique capability of network science to highlight interactions and connectivities among a large group of elements, it offers a refreshing perspective to analyze some of the most challenging problems in fluid dynamics in which interactions play important roles.  We are encouraged by the ongoing network-based research efforts from the fluid mechanics research community and look forward to the emergence of innovative approaches to understanding the connectivities that makes fluid mechanics vibrant.

\section*{Acknowledgements}

KT is grateful for the support from 
the US Army Research Office 
    (Grant: W911NF-19-1-0032 and W911NF-14-1-0386; 
    Program Managers: Matthew Munson and Samuel Stanton), 
the Air Force Office of Scientific Research 
    (Grants: FA9550-21-1-0178 and FA9550-16-1-0650; 
    Program Managers: Gregg Abate and Douglas Smith), 
the National Science Foundation 
    (Grant: 1632003; 
    Program Manager: Dimitrios Papavassiliou), 
and 
the Office of Naval Research 
    (Grant: N00014-19-1-2460; 
    Program Managers: Brian Holm-Hansen and David Gonazalez).  
AGN thanks the support from 
the National Science Foundation 
    (Grant: 2112085; Program Manager: Shahab Shojaei-Zadeh).
We also thank 
Shervin Bagheri,
Steven Brunton,
Eurika Kaiser, 
J.~Nathan Kutz, 
Vedasri Godavarthi, 
Melissa Green, 
Muralikrishnan Gopalakrishnan Meena, 
Bernd Noack, 
William Oates, 
David Rival, 
Taraneh Sayadi, 
Peter Schmid, 
R. I. Sujith, 
and
Chi-An Yeh 
for many enlightening discussions on the use of network science for fluid dynamics over countless cups of premium coffee.

\bibliographystyle{cas-model2-names}
\bibliography{refs_all,network_fluids}
\end{document}